\documentclass[a4paper,man,floatsintext,natbib]{apa6}

\usepackage[english]{babel}
\usepackage[utf8x]{inputenc}
\usepackage{amsmath}
\usepackage{graphicx}
\usepackage[colorinlistoftodos]{todonotes}

\usepackage{xcolor}
\usepackage{lipsum}  
\usepackage[lined,algoruled,commentsnumbered, linesnumbered]{algorithm2e}
\usepackage[normalem]{ulem}
\usepackage {tikz}
\usetikzlibrary{shapes,arrows}
\usetikzlibrary {positioning}
\usepackage{verbatim}
\definecolor {processblack}{cmyk}{0.96,0,0,0}
\usepackage{multirow}
\usepackage{amssymb,amsthm,amsmath,dsfont}
\tikzstyle{decision} = [diamond, draw, fill=green!40, 
    text width=8.5em, text badly centered, node distance=3cm, inner sep=0pt]
\tikzstyle{block} = [rectangle, draw, fill=red!20, 
    text centered, rounded corners, minimum height=0.1cm]
 \tikzstyle{node} = [circle, draw, fill=red!20, text width=0.2cm,
    text centered, rounded corners, minimum height=0.1cm]
\tikzstyle{line} = [draw, -latex']
\tikzstyle{cloud} = [draw, ellipse,fill=red!20, node distance=3cm,    minimum height=1em]

\usepackage{graphicx,subfigure}
\usepackage{soul}
\usepackage{nameref}
\usepackage{setspace}
\usepackage{hyperref}

\title{An analysis of community structure in Brazilian political topic-based Twitter networks}
\shorttitle{Community structure in political Twitter networks}
\author{Camila P.S. Tautenhain, Rodrigo Francisquini and Mariá C.V. Nascimento}
\affiliation{Instituto de Ciência e Tecnologia, Universidade Federal de São Paulo - UNIFESP, Av. Cesare Lattes, 1201, CEP: 12247-014 São José dos Campos, SP, Brazil}
\authornote{Corresponding author: Camila P.S. Tautenhain\\Email addresses: santos.camila@unifesp.br (Camila P.S. Tautenhain), francisquini.r@gmail.com (Rodrigo Francisquini) and  mcv.nascimento@unifesp.br (Mariá C.V. Nascimento)}

\abstract{
\onehalfspacing
Online social networks such as Twitter are important platforms for spreading public opinion on a variety of subjects.  The classification of users through the analysis of their posts on Twitter according to their opinion sharing can help marketing ads and political campaigns to focus on specific user groups.  Community detection-based techniques  are specially useful to classify Twitter users, as they do not require rule-based methods or labeled data to perform the clustering task. In this paper, we constructed networks using data related to political discussions in Brazil extracted from Twitter. We show that (i) these networks follow the power-law distribution, indicating that a few popular users are responsible for most of the ``mentions'' and ``retweets''; (ii) the most popular tweets are viral and spread across the communities whereas most of the remaining tweets are trapped in the communities where they originated; and (iii) words associated with positive sentiments are predominant in network communities related to the Brazilian presidential elections and appear in viral tweets.}


\begin{document}
\maketitle

\onehalfspacing

\section{Introduction}

Social networks have an important role in spreading {the political} views of their users{, who may be} public figures {or} common citizens \citep{Weller2014}.
Twitter, in particular, is an online social network {used to} post short texts, called \textit{tweets}, through which {the users} express {their opinions}, thoughts and feelings. 
In addition, users can make or share public statements {they agree with} \citep{pak2010twitter}. {They may} also interact with each other on Twitter {through} the option ``follow'' and by retweeting {other tweets}. In this sense, Twitter presents valuable information {on} public opinion on various areas and can be employed for a variety of purposes, {e.g.}, to direct marketing ads \citep{cambria2013new} and to guide political campaigns \citep{larsson2012studying,Golbeck2010,Yardi2010}. This online social network has approximately  321 million active users \citep{verge}. Brazil is one of the 10 countries with the highest number of active accounts in Twitter according to information provided by the company. Therefore, {data mining algorithms play a key role in} the  analysis of networks created {with} data collected from Twitter.

{Context mining, in particular,} is a research topic that has attracted {the} attention of the scientific community  \citep{Conover2011_predicting,Pennacchiotti2011_democrats,Makazhanov2014,Surian2016,Gaul2017,Ahmadi2018}.
A number of tools has been proposed for {the} analysis {of} Twitter data, among {which} we highlight {the ones} based on  natural language processing techniques \citep{ding2007utility,Sun2017} and supervised machine learning algorithms \citep{Pennacchiotti2011_democrats,Conover2011_predicting,Makazhanov2014}. 
These strategies aim at  detecting {users' opinions and feelings according to the content of what they publish.}
A major drawback {of} natural language processing techniques is the diversity of linguistic rules. Furthermore, supervised machine learning methods can {only be used}  when there is enough labeled data {to train} the algorithm.

Graph-based techniques, on the other hand, compile Twitter data into graphs {where} the vertices are the users and the arcs {are the} interactions between users. These techniques use graphs with unlabeled data to detect public opinions. Examples of graph-based techniques can be found in \citep{Conover2011_predicting,Pennacchiotti2011_democrats,Sluban2015,Surian2016}.
According to a social network property known as homophily \citep{Tang2014}, users are more likely to retweet or mention tweets {from other} users {they share a common interest with}. {Community} detection in networks, thereby, can provide useful information {on} whether or not {tweets share} the same contexts.

In light of the recent corruption investigations in Brazil and the presidential elections, {discussions} in social networks about the Brazilian political scenario {have seen a significant increase}.
{In the study reported here}, we extracted data {from tweets} containing words related to politicians who would reportedly run for the 2018 Brazilian presidential elections and to corruption {investigations} in Brazil. The Twitter API \citep{twitterapi} was employed to extract data concerning {interactions between users}, such as retweets and mentions, to construct the Twitter networks proposed in this paper.

The primary purpose of this {study was} to analyze the structure and topology of networks constructed {with data from Twitter in order to classify} users according to the subject of their tweets using community detection algorithms and to investigate the sentiment  associated with the communities in the networks related to the Brazilian {presidential} elections.
We also confirmed the observation {by \citet{Weng2013}} that a few popular tweets, referred to as viral tweets, spread across the communities like viruses and, therefore, are not expected to affect the community structure.
{Non-viral} tweets, on the other hand, get trapped in the communities {where} they originated and, thereby, are expected to {stay} within the communities.

The networks under investigation are scale-free {and therefore have} few {high-degree vertices} and a large number of {low-degree} vertices.  In other words, a few popular users contribute to most of the retweets and mentions on these networks.  Thereby, the {most retweeted tweets are the ones} posted by the most popular users. In addition, a case study {on the} networks related to the Brazilian {presidential} elections showed that {words} associated with positive sentiments are predominant in the communities. Amongst these words, the ones which contribute {most} to the positive sentiment in the communities are the {same words} that appear in viral tweets.

The {remainder} of {the} paper is organized as follows: Section \nameref{sec:rev_community_detection} introduces basic concepts and algorithms for community detection in networks; Section \nameref{sec:related_works} presents a brief discussion about related works; Section \nameref{sec:database} shows the networks constructed from Twitter data;  Section \nameref{sec:computational_experiments} presents the computational experiments carried out with community detection algorithms; 
Section \nameref{sec:network_analysis} presents a network analysis where the structure of the introduced networks is investigated; Section \nameref{sec:community_classification} investigates the communities related to the 2018 Brazilian {presidential} elections and the sentiments associated with them;
and Section \nameref{sec:conclusion} summarizes the contributions of the paper and gives directions to {future} works.

\section{Community Detection in  Networks}\label{sec:rev_community_detection}

This section presents {the} theoretical background and {primary} notations used throughout the paper. Moreover, it introduces some of the classical community detection algorithms employed in studies related to our investigation found in the literature{, and which} will also be the tools {used for} the network analysis in this paper.

\subsection{Basic Concepts and Notations}\label{sec:notations}

In this paper, a network is defined as a directed graph (digraph)  $G=(V,A,\omega)$, where $V$ and $A$ are, respectively, its sets of vertices and arcs  and  $\omega:A \rightarrow \mathds{N}$ is a function that assigns a weight to each arc of the digraph. An arc $e$ is defined {as} an ordered tuple $(v,u)$, where {the points} $v,u \in V$  are  called  tail and head of $e$, respectively. Let $n=|V|$, $m=|A|$ and $\Omega= \sum_{e \in A}{\omega(e)}$ be, respectively, the number of vertices, number of arcs and the total weight of arcs of $G$.

The in- and out-degrees of a vertex $v$ are given by, respectively, $d^{in}(v)=\sum_{(z,v)\in A} \omega(z,v)$ and  $d^{out}({v})=\sum_{(v,z)\in A} \omega(v,z)$. Moreover, the degree of a vertex $v$ is defined by ${d(}v)=d^{in}(v)+d^{out}(v)$.

A partition of $V$ into $k$ communities is given by  $\mathcal{P}=\{P_1,P_2,\dots,P_k\}$, where  $\bigcup_{i=1}^k P_i=V$ and $P_i \cap P_j=\emptyset$, $\forall  i,j \in \{1,2,\dots,k \}, i\neq j$. The community of a vertex $v$ is referred to as $\mathcal{P}(v)$.
Moreover, let $G[V']$ be the vertex-induced subgraph, which is the subgraph of $G$ such that $V(G[V']) =V'$ and $A(G[V'])$ is composed by all arcs in $A$ {whose} points are in $V'$.

\subsection{Community Evaluation Metrics}

One way of evaluating the quality of the community structure of a given partition is by calculating its closeness to the expected partition according to {metrics}, like the Normalized Mutual Information (NMI) proposed in \citep{Shannon1948}. However, as the expected partition is not available in most applications, {in order to evaluate the quality of partitions}, we employed the widely used evaluation measure known as modularity.

Modularity is based on the intuitive notion that vertices belonging to the same community are more likely to connect with each other than with vertices from different communities.
Equation \eqref{eq:modularity} defines the modularity measure for weighted undirected graphs \citep{Leicht2008} {applied to a partition $\mathcal{P}$ {of} digraph $G$}.

 \begin{equation}
Q(\mathcal{P}) = \frac{1}{2 \Omega} \sum_{P \in \mathcal{P}} \sum_{{(i,j)\in A(G[P])}}\left( \omega(i,j) -  \frac{d(i) d(j)}{4 \Omega} \right)  
  \label{eq:modularity}
 \end{equation}

{As can be seen in} Equation \eqref{eq:modularity}, the modularity of $\mathcal{P}$ consists {in} the  difference between the total weight of the arcs between vertices of a same community {and} the expected sum of weights of edges inside communities in a random undirected graph whose vertices have the same degree sequence as the digraph under consideration. The values of the measure range from $-1$ to $1$ and the higher the value, the better the partition.

\subsection{Community Detection Algorithms}

Since the modularity measure {was proposed by} {\citet{Newman2004c}},  several authors {have studied ways} to detect {community structures} in a plethora of complex networks. 
In particular, community detection algorithms {that} aim at maximizing the modularity measure have been extensively studied  \citep{Newman2004,Blondel2008,Santos2016a,Francisquini2017}.
\citet{Blondel2008} presented the Louvain method, a multilevel greedy algorithm {for} modularity maximization  that was successfully applied to a web graph with 118 million vertices and 1 billion edges. \citet{Newman2006} studied a {spectral} decomposition {using} the leading eigenvector of the modularity matrix  and introduced a greedy algorithm to define the communities.

Despite the good results {achieved} by modularity maximization-based methods, the measure has a resolution limit that {restricts} the size of the communities found by the algorithms. \citet{Yang2016} showed that the leading eigenvector algorithm {proposed by} \citet{Newman2006} found partitions far from the expected even in networks with small mixture coefficients\footnote{The mixture coefficient of a network is the percentage of edges or arcs crossing communities with respect to the total number of edges or arcs in the graph.}. \citet{Yang2016} also showed that the Louvain method was able to find the expected communities in networks with small mixture coefficients {and partitions sufficiently close to the ones expected for the networks} with high mixture coefficients.

Among the algorithms that aim {to optimize} different measures, we highlight those proposed in \citep{Pons2005,Rosvall2007,Lancichinetti2011_oslom}.
\citet{Pons2005} introduced the Walktrap algorithm to detect  vertex partitions {based on the random walkers' probabilities of finding trails between vertices in the networks}.
 \citet{Rosvall2007} introduced the map equation measure to calculate the description length of a random walker in a digraph. The Infomap algorithm, also proposed by the authors, minimizes the map equation to find communiries in networks.
Walktrap achieved results comparable {with the ones achieved by} Louvain, whereas Infomap {achieved} worse {results in} networks with higher mixture coefficients \citep{Yang2016}.

\citet{Lancichinetti2011_oslom} suggested {finding} communities in networks by evaluating the communities {according to} their  statistical significance. They proposed an algorithm named Order Statistics Local Optimization Method (OSLOM), {which obtained} better results than all the aforementioned algorithms in networks with the higher mixture coefficients, according to \citet{Lancichinetti2011_oslom}. Nevertheless, OSLOM requires more computational time than the other algorithms to find the communities.

The  community detection algorithm known as Label Propagation (LP)  \citep{Raghavan2007} is neighborhood-based and does not optimize any measure. To form the communities, it starts from a partition whose vertices are isolated in their own communities -- singletons --- and iteratively assigns each vertex to the community {the} majority of its neighbors belong to. The computational complexity of LP is  quasi-linear, being very fast in detecting communities. However, {LP} is highly dependent on the number of iterations since if it is too high, it tends to merge all communities into a single cluster.

Next section presents a brief literature review of studies related to the classification of users in Twitter networks.

\section{Related Works}\label{sec:related_works}

This section {briefly} discusses classical approaches for mining public opinion in social networks through natural language processing and {s}upervised machine learning algorithms. Special attention is given to applications using  information extracted from Twitter  and to  works that classify users based on community detection algorithms.

Natural language processing techniques use a set of  words that describe opinion and linguistic rules to infer {users' opinions} on Twitter. 
Although rule-based techniques are rather effective, linguistic rules usually change along the years and differ from language to language. In addition, users always follow  linguistic rules in their posts. Thus, these techniques are very limited and {strongly} dependent {on} specialist knowledge \citep{wang2011topic}. For further details on these techniques, we refer the reader to the recent review in \citep{Sun2017}.

Supervised machine learning approaches are commonly applied to different domains and {can therefore} overcome the challenges {presented by} rule-based methods. Among these approaches, we highlight the methods studied in \citep{Conover2011_predicting,Pennacchiotti2011_democrats,Makazhanov2014}. \citet{Pennacchiotti2011_democrats} trained a decision tree to learn Twitter users' information, such {as} political affiliation, ethnicity and whether {the users were} fans of a famous coffee shop {according to} the number of followers and friends, number of replies, average number of tweets and linguistic content of tweets. \citet{Makazhanov2014} also trained a decision tree coupled with a logistic regression to learn about political positions through the {interaction} between users. 
\citet{Conover2011_predicting} predicted the political alignment of Twitter users {by using} a Support Vector Machine (SVM) trained with the contents of users' tweets.

However, these supervised methods require labeled data to be used as training data. Generally, real-time data are not inherently labeled.  In such cases, a specialist can manually label a small sample for  training machine learning algorithms, as performed in \citep{Conover2011_predicting}. 
In dealing with big data, manual data labeling is not viable, since if a very small sample is used for training, the model becomes unreliable \citep{Raudys1991}.

Community detection in networks {does} not need any  prior knowledge about the data classes and mostly {does} not rely on rule-based techniques. These algorithms detect communities whose users interact more {among} each other, in terms of friendship, retweets or mentions, than with the rest of the users. {Community} detection in networks {can thereby} provide useful sharing information {on} tweets among {users} \citep{Conover2011_predicting,Sluban2015}. 
In addition to the SVM classifier, \citet{Conover2011_predicting} employed a community detection algorithm based on the leading eigenvector method \citep{Newman2006} and on the label propagation algorithm \citep{Raghavan2007}. 
\citet{Sluban2015} detected the communities of a Twitter network using the Louvain method \citep{Blondel2008} and then identified the topics of the tweets in each  community using a content-based classification algorithm.

 \citet{Surian2016} investigated discussions about HPV vaccines {on} Twitter using a community detection algorithm to infer about the topics of the tweets. 
Despite not explicitly employing a community detection algorithm, \citet{Pennacchiotti2011_democrats}  refined the results found by a supervised classification method by updating the classes of the users according to the classes {their} friends {belonged} to.

\citet{Weng2013} studied the spreading of hashtags on Twitter networks and {observed} that a few viral hashtags {disseminate through} the communities like viruses, while {most} get trapped within the communities where they originated.
According to \citet{Weng2013}, {the virality of a tweet can be measured {as the} percentage of retweets it receives from users of {communities different} than the community of the user who posted it.}
Let $\mathcal{P}$ be a partition of vertices into communities. Equation \eqref{eq:virality} defines the virality of a tweet $tw$ posted by a user $v$ from a community $\mathcal{P}(v)$.

\begin{equation}
    vir(tw) = 
    \frac{\sum_{j \in \mathcal{J}, \mathcal{P}(j)\neq \mathcal{P}(v)}  {\omega}({v,j})}
    {\sum_{j \in \mathcal{J}} {\omega}({v,j})}
    \label{eq:virality}
\end{equation}

\noindent where $\mathcal{J}=\{j \in V \mbox{ such that } (v,j) \in A \mbox{ and } j \mbox{ retweeted } tw \}$ is the set of users who retweeted tweet $tw$.

As \citet{Weng2013} affirmed that viral tweets spread equally to the communities, we can expect that {they} do not affect the community structure of the networks. Non-viral tweets, on the other hand, spread mostly {in} the {communities}  where they {originated} and, thus, are expected to define the community structure of the networks.

 Next section  discusses the {methodology} employed to extract information from Twitter to construct the networks under evaluation in this paper.

\section{Extracted Networks}\label{sec:database}

In this paper, we constructed directed networks {from} data collected {on} Twitter using the streaming API {\citep{twitterapi}}{, which} returns {samples of} recent tweets {matching the application queries.} 
 The vertices of the digraphs represent the Twitter users from the sample of tweets. {Arc} $e=(v,u)$ {in} digraph $G$ expresses that user $v$ was  mentioned, {replied to} or {posted} a tweet $tw$ that was retweeted by user $u$. The weight {of} arc $e$, ${\omega(e)}$, is the number of {times} $v$ was mentioned, {replied to} or {retweeted} by user $u$. 

\begin{figure}[!htb]
\begin{center}
\small
\begin{tikzpicture}[node distance = 0.6cm, auto]
    \node [node] (v3) {$z$};
    \node [node, right=3.5cm of v3] (v1) {$v$};
    \node [node, right=3.5cm of v1] (v2) {$u$};
    \path[line] (v1) --node{$v$ was retweeted by $u$ } (v2);
    \path[line, above=2cm] (v1) --node{$v$ was mentioned by $z$ } (v3);
\end{tikzpicture}
\caption{Example of a network with 3 users and whose arcs were created due to retweets and mentions.}
\label{fig_ex_arc_tweet}
\end{center}
\end{figure}
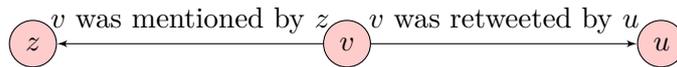

Figure \ref{fig_ex_arc_tweet} shows a triplet whose arcs were added due to tweets. In this figure,  user $u$ retweeted a tweet {from} user $v$, creating arc ($v,u$). In addition, user $z$ mentioned user $v$ in another tweet, represented by arc $(v,z)$.  Note that the arcs of the networks do not {distinguish retweets from mentions}.
 
{Table \ref{tab:network_queries} presents the queries employed to construct the networks. Table \ref{tab:network_network} presents the number of vertices ($n$) and arcs ($m$); total weight of the arcs ($\Omega$){; and} the average (avg), standard deviation (std), minimum (min) and maximum (max) in- and out-degrees {of the} proposed networks.}
Note that the sum of the in-degrees and the sum of the out-degrees of all vertices are equal to the total weight of the arcs, $\Omega$, since each arc $e=(v,u)$ of $G$ counts $\omega{{(e)}}$ to the  out-degree of $v$ and to the in-degree of $u$. Consequently, the average in- and out-degrees are the same. The networks introduced in this paper have the maximum out-degree  at least $13.55$ times larger than the maximum in-degree.

{The \textit{Politicians} network} was constructed by querying Twitter API tweets {matching} the names of politicians who {demonstrated} interest in running {for the} 2018 Brazilian {presidential} elections. As this network was constructed before the politicians {had effectively} presented {their} candidacy, we selected the names based on an article published {by the} BBC \citep{bbc_article_election}. 
{The \textit{Bolsonaro} and \textit{Lula} networks}
 captured {interactions in} tweets that {referred} to politicians Jair Bolsonaro and Luiz Inácio Lula da Silva, respectively; {\textit{LavaJato} network was} formed by tweets presenting reactions {to} Operation Car {Wash, in Brazil, which} investigates executives {who have} reportedly  accepted bribes from construction firms in exchange for inflating contracts prices.
{\textit{1st\_Round} and \textit{2nd\_Round} networks} were constructed using, respectively, tweets expressing reactions to the first and second rounds of the 2018 Brazilian {presidential} elections\footnote{The Brazilian {presidential} elections {are} composed {of} two rounds. In case the most voted candidate in the first round does not get the absolute majority of the votes, the second round of the election occurs. In the second round, the two most {voted candidates} in the first round are the only {ones} to run.}. These last two networks only model interactions using retweets.

\begin{table}[htb]
    \centering
    \caption{Information on the queries submitted to Twitter streaming API.}
    \small
    \label{tab:network_queries}
    \setlength{\tabcolsep}{1pt}
    \begin{tabular}{|c|c|c|}
    \hline
        \multirow{1}{*}{Network}              &   \multirow{1}{*}{Query}  & \multirow{1}{*}{Date} \\ \hline
     
        \multirow{3}{*}{\textit{Politicians}}
        & lula,bolsonaro,alckmin,temer,cirogomes,  marinasilva,      
        &  \multirow{2}{*}{May 15, 2018 to} \\ 
        &  alvarodias,  fernandohaddad, manueladavila,guilhermeafif,  & \multirow{2}{*}{May 16, 2018} \\  
        &  fernandocollor, guilhermeboulos,joaoamoedo  &  \\  \hline

        \multirow{2}{*}{ \textit{Lula}}
        & lulapreso,lula preso,lulalivre,lula livre,   & May 7, 2018 to\\
       &  & May 9, 2018 \\  \hline

       \multirow{2}{*}{\textit{Bolsonaro}}
       & bolsonaro, bolsonaro2018, jair bolsonaro,  bolsonaro presidente, &   May, 7 2018 to\\ 
       & \#bolsonaro2018, \#bolsonaropresidente, \#bolsonaro    &     May 14, 2018 \\   \hline
         \multirow{2}{*}{\textit{LavaJato}}  
        &  \multirow{2}{*}{lavajato,lava-jato,``lava jato''}  &  May, 7 2018 to
        \\
         &    &  May 14, 2018\\
        \hline
   \multirow{4}{*}{\textit{1st\_Round}}
        & jairbolsonaro,bolsonaro,fernandohaddad,haddad, cirogomes, 
        &   \multirow{4}{*}{7 October, 2018} \\ 
        & ciro,geraldoalckmin,alckmin, marinasilva, marina   & \\  
        &  alvarodias,henriquemeirelles, meirelles, joaoamoedo,amoedo&\\  
        & guilhermeboulos, boulos,  josemariaeymael,cabodaciolo  &\\   \hline
        \multirow{4}{*}{\textit{2nd\_Round}}
        & jairbolsonaro,bolsonaro,fernandohaddad,haddad, cirogomes, &   \multirow{4}{*}{28 October, 2018}\\ 
        & ciro,geraldoalckmin,alckmin,marinasilva,marina,  &   \\  
        &  alvarodias,henriquemeirelles, meirelles, joaoamoedo,amoedo     &\\  
        & guilhermeboulos, boulos,  josemariaeymael,cabodaciolo      &\\   \hline
 
    \end{tabular}
\end{table}

\begin{table}[htb]
    \centering
    \caption{Information on the networks extracted from Twitter.}
    \small
    \label{tab:network_network}
    \setlength{\tabcolsep}{2pt}
    \begin{tabular}{|c|c|c|c|c|c|c|c|c|c|c|c|}
    \hline
        \multirow{2}{*}{Network}                &  \multirow{2}{*}{$n$} &  \multirow{2}{*}{$m$} &  \multirow{2}{*}{$\Omega$} &    \multicolumn{4}{|c|}{in-degree} & \multicolumn{4}{|c|}{out-degree}  \\ \cline{5-12}
            &   & &   & avg & std & min & max & avg & std & min & max   \\ \hline
     
        \multirow{1}{*}{\textit{Politicians}}
        
        & \multirow{1}{*}{16137} & \multirow{1}{*}{31933} & \multirow{1}{*}{40858} & \multirow{1}{*}{2.53} & \multirow{1}{*}{6.06} & \multirow{1}{*}{0} & \multirow{1}{*}{359} & \multirow{1}{*}{ 2.53} & \multirow{1}{*}{32.39 } & \multirow{1}{*}{0} & \multirow{1}{*}{2875}  \\   \hline

        \multirow{1}{*}{ \textit{Lula}}
                            & \multirow{1}{*}{13319} & \multirow{1}{*}{33519}  &  \multirow{1}{*}{40456}
        & \multirow{1}{*}{3.04} & \multirow{1}{*}{7.8} & \multirow{1}{*}{0} & \multirow{1}{*}{236} & \multirow{1}{*}{3.04} & \multirow{1}{*}{35.8} & \multirow{1}{*}{0} & \multirow{1}{*}{1798} \\
                           \hline

       \multirow{1}{*}{\textit{Bolsonaro}}
       
        & \multirow{1}{*}{873} & \multirow{1}{*}{1264} & \multirow{1}{*}{1676} & \multirow{1}{*}{1.92} & \multirow{1}{*}{2.75} & \multirow{1}{*}{0} & \multirow{1}{*}{32} & \multirow{1}{*}{1.92} & \multirow{1}{*}{9.62} & \multirow{1}{*}{0} & \multirow{1}{*}{178}  \\ 
          \hline
         \multirow{1}{*}{\textit{LavaJato}}  
                      &   \multirow{1}{*}{14081} &  \multirow{1}{*}{35642}
        &  \multirow{1}{*}{42937} & \multirow{1}{*}{3.05} & \multirow{1}{*}{7.8} & \multirow{1}{*}{0} & \multirow{1}{*}{237} & \multirow{1}{*}{3.05} & \multirow{1}{*}{35.16} & \multirow{1}{*}{0} & \multirow{1}{*}{1798}  
        \\        \hline
   \multirow{1}{*}{\textit{1st\_Round}}
        & \multirow{1}{*}{135865} & \multirow{1}{*}{242679} & \multirow{1}{*}{249042} & \multirow{1}{*}{1.83} & \multirow{1}{*}{2.49 } & \multirow{1}{*}{ 0} & \multirow{1}{*}{103} & \multirow{1}{*}{1.83} & \multirow{1}{*}{42.76} & \multirow{1}{*}{ 0 }  &   \multirow{1}{*}{6032}  \\   \hline
        \multirow{1}{*}{\textit{2nd\_Round}}
       
        & \multirow{1}{*}{100755} & \multirow{1}{*}{171813} & \multirow{1}{*}{176955} & \multirow{1}{*}{1.76} & \multirow{1}{*}{2.38} & \multirow{1}{*}{ 0} & \multirow{1}{*}{79} & \multirow{1}{*}{ 1.76 } & \multirow{1}{*}{33.17} & \multirow{1}{*}{ 0 } & \multirow{1}{*}{3863}  \\  \hline
 
    \end{tabular}
\end{table}
The following section presents computational experiments with community detection algorithms {in these networks.}

\section{Computational Experiments with Community Detection Algorithms}\label{sec:computational_experiments}

In {the experiments presented in this section}, we considered four different community detection algorithms available in the igraph package \citep{igraph}: Louvain Method \citep{Blondel2008}, Label Propagation (LP) \citep{Raghavan2007},  Infomap \citep{Rosvall2007} and Walktrap \citep{Pons2005}.
Even {though the} Infomap {method} has a version to detect communities in digraphs, the version to undirected graphs found communities with higher modularity values and was therefore selected for the experiments. All the experiments were carried out on a computer with {an} Intel Core Xeon E5-1620 processor with 3.7-GHz and 32GB of main memory.

Table \ref{tab:modularity} shows the modularity of the partitions obtained by the algorithms {for} each of the networks used. In this table, the columns marked as ``Modularity'' refer to the {modularity value} of the partitions obtained by the community detection {algorithms according to Equation \eqref{eq:modularity}.}
The columns marked as ``Time'' report the time (in seconds) required to detect the communities. Due {to} computer memory limitations, Walktrap could not find the communities for networks \textit{1st\_Round} and \textit{2nd\_Round}.

\begin{table}[htb]
    \centering
    \small
    \caption{Modularity values {of communities} achieved by reference community detection algorithms.}
    \label{tab:modularity}
    \setlength{\tabcolsep}{2pt}
    \begin{tabular}{|c|c|c|c|c|c|c|c|c|}
    \hline
        \multicolumn{1}{|c|}{\multirow{2}{*}{Network}}              & \multicolumn{2}{|c|}{Louvain}   &  \multicolumn{2}{|c|}{LP}   & \multicolumn{2}{|c|}{Infomap}        & \multicolumn{2}{|c|}{Walktrap}  \\  \cline{2-9}
                        & Modularity  & Time & Modularity &  Time & Modularity &  Time & Modularity  & Time \\ \hline
\textit{Politicians} & 0.7211 &  0.16 & 0.6521 & 0.2 & 0.6162 & 10.09 & 0.6527 &  8.98 \\ \hline
\textit{Lula} & 0.5464 & 0.15 & 0.4107 & 0.14 & 0.4676 & 8.77 & 0.4465 &  7.92 \\ \hline
\textit{Bolsonaro} & 0.7993 & 0.01 & 0.746 & 0.01 & 0.7497 & 0.15 & 0.7435 & 0.02 \\ \hline
\textit{LavaJato} & 0.5504 &  0.17 & 0.4247 & 0.15 & 0.4673 & 9.58 & 0.4594 &  8.89 \\ \hline
\textit{1st\_Round} & 0.6771 & 1.48 & 0.6488 & 8.48 & 0.5401 &  2.98 & - & -  \\ \hline
\textit{2nd\_Round} & 0.7228 &  1.54 & 0.6664 & 2.52 & 0.5742 &  1.95 & -  & - \\ \hline
\end{tabular}
\end{table}

The Louvain method is the only algorithm included in Table \ref{tab:modularity} {that} aims at maximizing the modularity. Thereby, as expected, it achieved the highest modularity values for all the networks.  The remaining algorithms, however, still obtained partitions with high modularity values. As previously explained, the modularity measure was selected to evaluate the quality of the partitions because there are no ground-truth communities available for these constructed networks. Therefore, we selected the Louvain method to detect communities in networks in the {remainder} of this paper {as it was} the algorithm that obtained communities with the highest modularity values in the lowest computational times.

\begin{figure}[!htb]
\begin{center}
 \subfigure[fig:indegree][{\textit{1st\_Round} network}.]{\label{fig:stream_Elections_1st}\includegraphics[width=0.44\textwidth]{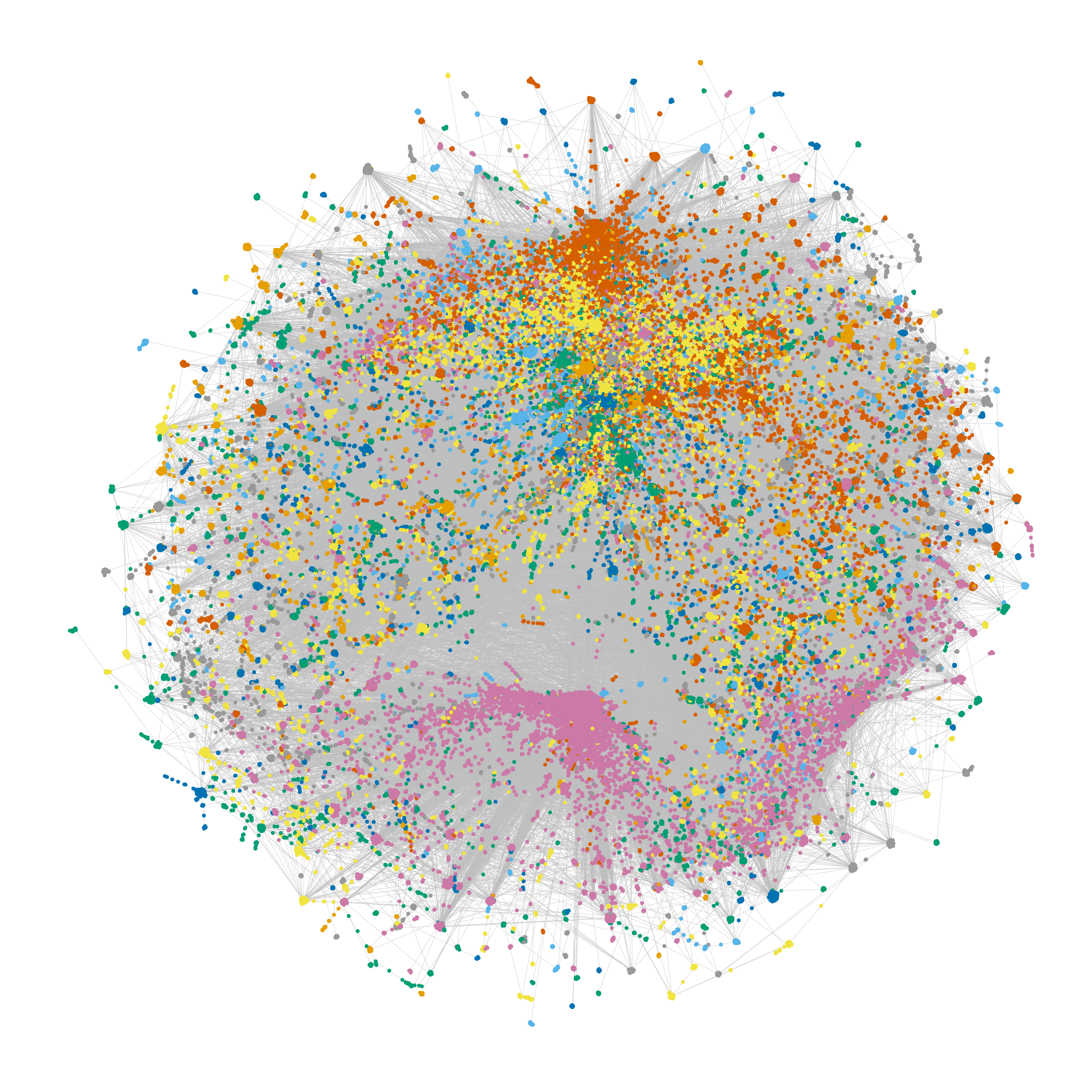}}
  \subfigure[fig:indegree][{\textit{2nd\_Round} network}.]{\label{fig:stream_Elections_2nd}\includegraphics[width=0.44\textwidth]{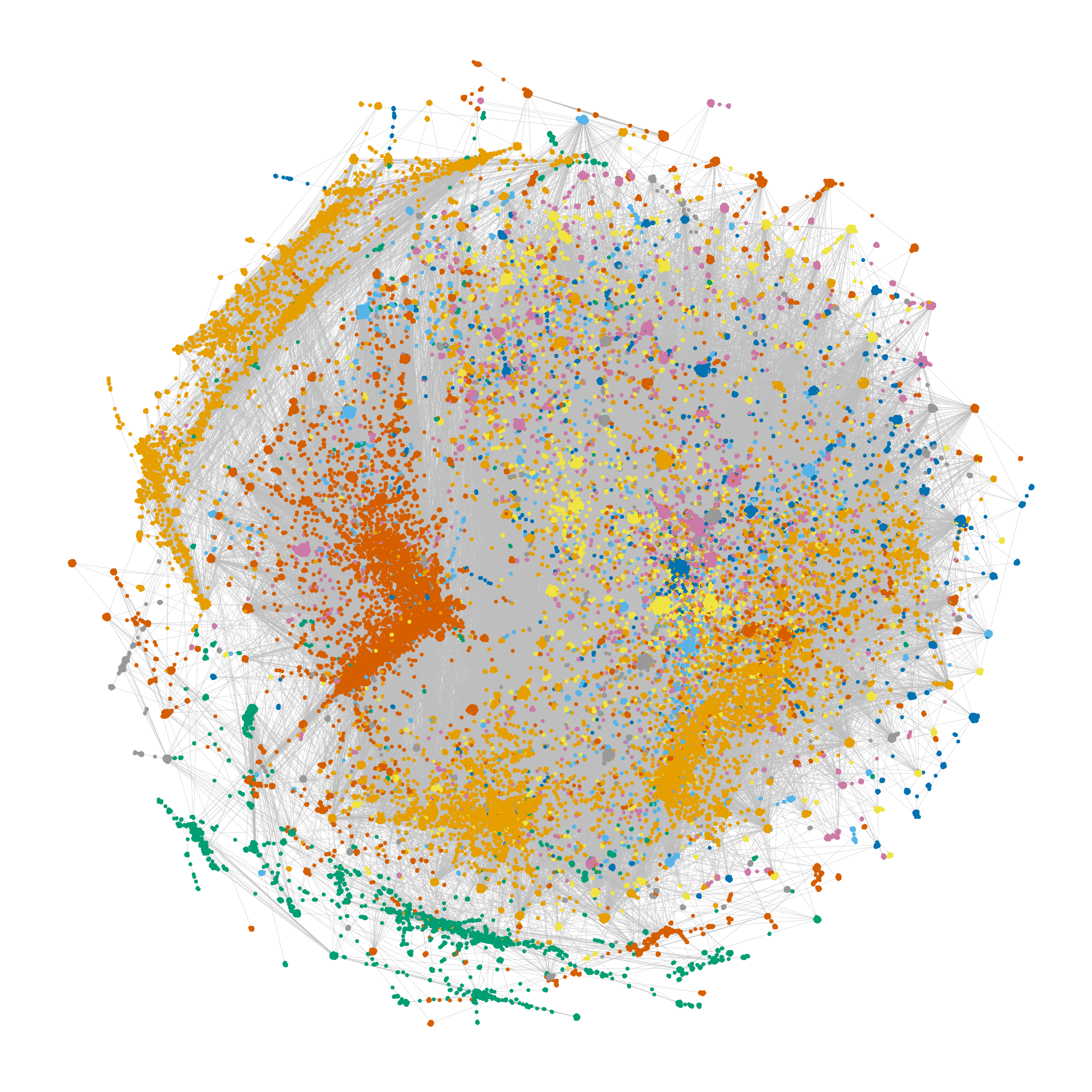}}
  \caption{Communities found by the Louvain method.}
      \label{fig:graphs}
\end{center}
\end{figure}

The partitions achieved by the Louvain method {for} the  {\textit{1st\_Round} and \textit{2nd\_Round} networks} were chosen to be analyzed at length in this paper. To illustrate these partitions, Figures \ref{fig:stream_Elections_1st} and \ref{fig:stream_Elections_2nd} exhibit the {\textit{1st\_Round} and \textit{2nd\_Round} networks} along with the communities found. In these figures, vertices belonging to the same community have the same color. 

Next section presents {the} analysis performed to describe the topology and the community structure of the constructed networks.

\section{Network Analysis}\label{sec:network_analysis}

\citet{Tedjamulia2005} suggested {dividing} the users {of Twitter networks} into categories according to their popularity {and} activity. In this paper, {we have} divided the users into 8 sets (categories) {according} to their  popularity -- number of mentions and retweets received {by the user} -- and activity -- number of mentions and retweets {made} by {the} user. The categories related to the popularity are called $PC1$, $PC2$, $PC3$ and $PCR$, whereas those regarding activity are $AC1$, $AC2$, $AC3$ and $ACR$.
$PC1$ and $AC1$ contain the $0.1\%$ most popular and the $0.1\%$ most active users, respectively; $PC2$ and $AC2$ comprise the $1\%$ most popular users {not} in $PC1$ and the $1\%$ most active users {$AC1$}, respectively; $PC3$ and $AC3$ have the $10\%$ most popular users  that are neither in $PC1$ {nor in} $PC2$ and  the $10\%$ most active users  that are neither in $AC1$ {nor in} $AC2$; and $PCR$ and $ACR$ {comprise} as the $90\%$ least popular and the $90\%$ least {active users,} respectively.

According to these sets of users, the $1\%$ most popular and active users are those from $PC1\cup PC2$ and $AC1\cup AC2${, respectively}. The $10\%$ most popular and active users are those that belong to $PC1\cup PC2\cup PC3$ and $AC1\cup AC2 \cup AC3${, respectively}.

Now, consider the division of tweets into sets (categories) according to their popularity regarding only the number of retweets they received:
$TC1, TC2, TC3$ and $TCR$ are the categories of tweets {and are respectively ranked as the} $0.1\%$ most popular {tweets}, $1\%$ most popular tweets {not} in $TC1$,  $10\%$ most popular tweets {not in} $TC1$ nor in $TC2$ and  $90\%$ least popular tweets.

The following sections show the analysis of the constructed networks with respect to the popularity and activity of their {users, tweets and communities. Moreover, they also show the evolution of the topologies of the networks along time.} The classification of users into communities and the spread of viral tweets is also discussed. In particular, the results of the largest network \textit{1st\_Round} are further investigated.

\subsection{Users' Popularity and Activity}

 This section {presents} the analysis of the proposed networks with respect to {the distribution of their degrees in order to} describe the popularity and activity of the users {and} their influence on the constructed networks. {The section investigates the {\textit{1st\_Round}} network more thoroughly} whereas the  other networks are loosely discussed.

The popularity of a user is measured by the total number of mentions and retweets that they receive. A user's activity is defined by the number of mentions and retweets they make.
According to the definition of arcs presented in Section \nameref{sec:database}, the in- and out-degrees of a vertex represent, respectively, the activity and the popularity of the user that the vertex refers to.

\begin{figure}[!htb]
\begin{center}
\subfigure[fig:outdegree][Out-degree distribution.]{\label{fig:outdegree}\includegraphics[width=0.44\textwidth]{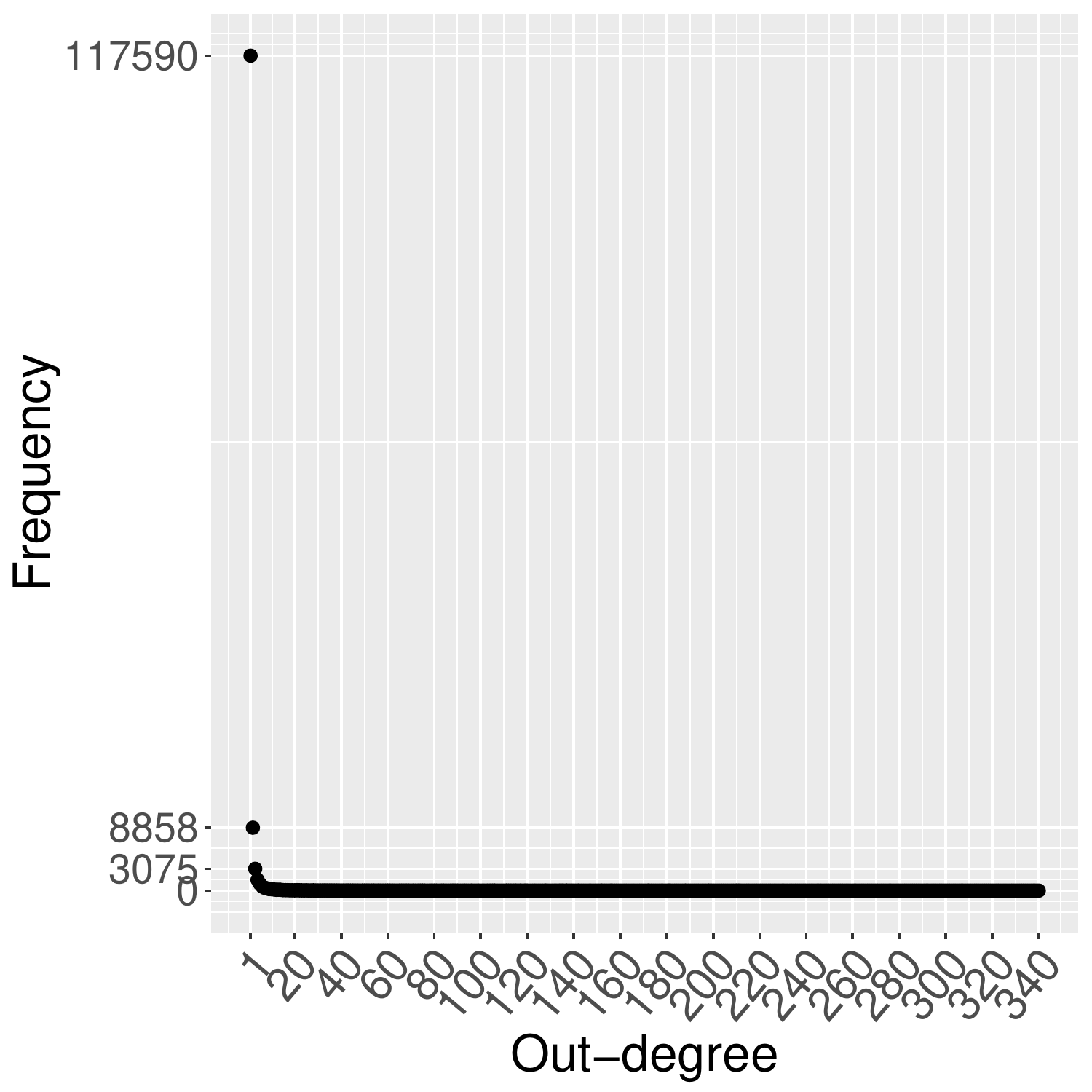}}
  \subfigure[fig:indegree][In-degree distribution.]{\label{fig:indegree}\includegraphics[width=0.44\textwidth]{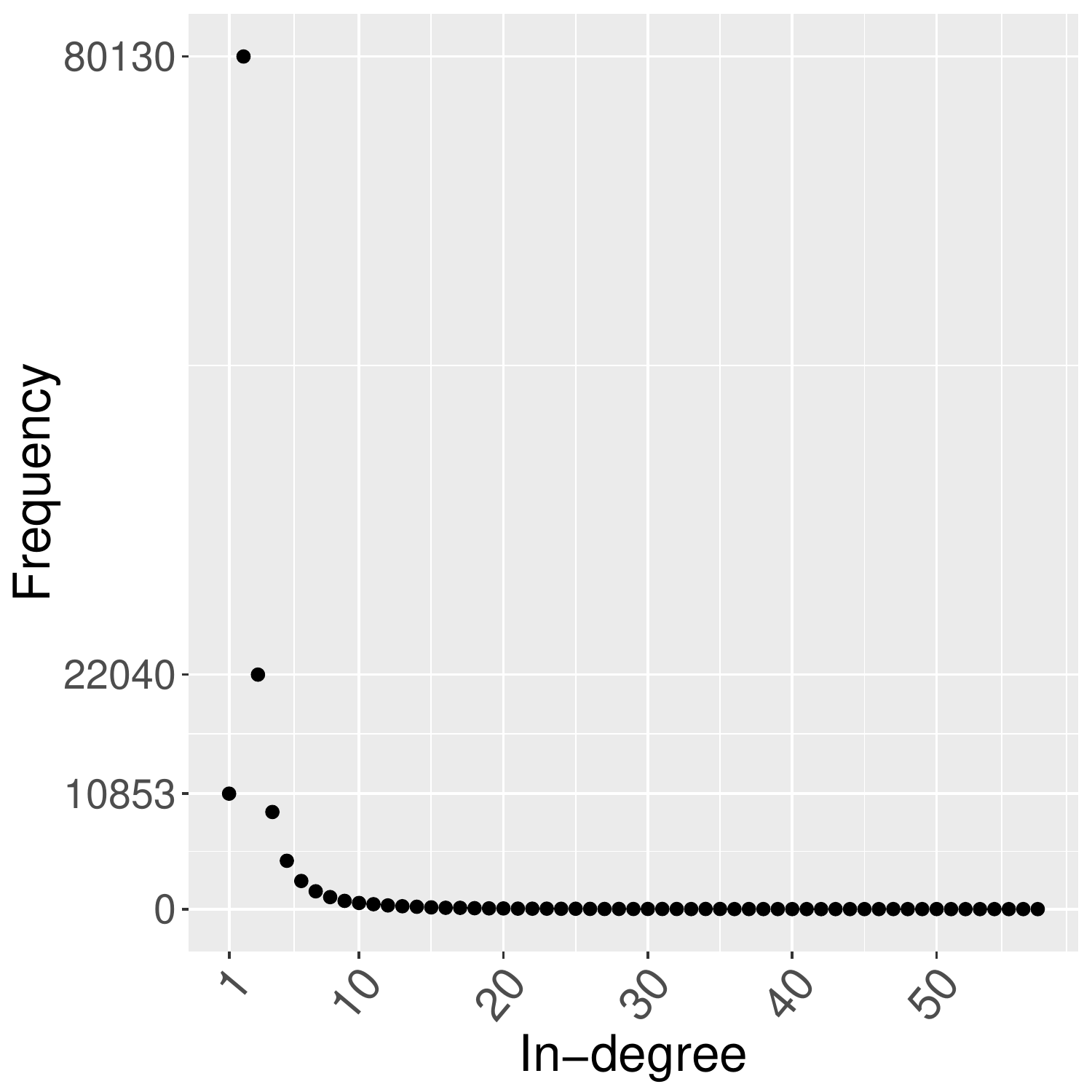}}
       \caption{Histograms of the out{-} and in-degree distributions of \textit{1st\_Round}.} 
      \label{fig:degree_distr}
\end{center}
\end{figure}

Large social networks are usually scale-free. In particular, on Twitter-based networks,  the most visible users  are the nodes with the highest degrees. To analyze the topology of the constructed networks, Figures \ref{fig:indegree} and \ref{fig:outdegree} exhibit histograms with the out- and in-degree frequencies of  {{\textit{1st\_Round}}}.

{As can be noted in} Figure \ref{fig:degree_distr}, the distribution of the node degrees follows {a} power law. The power law degree distribution of the vertices corroborates with the observation that a few users are more active or visible in social networks.

\citet{Bruns2013} stated that {a} minority of users is very active or popular in a Twitter network.
In the \textit{1st\_Round} network, users from $ACR, AC3, AC2$ and $AC1$ {were} responsible for, respectively, $61.08\%, 28.54\%, 8.28\%$ and $2.11\%$ of the total number of tweets. Users from $PCR, PC3, PC2$ and $PC1$ {received}, respectively, $1.88\%, 15.75\%, 29.62\%$ and $52.75\%$ of the total number of retweets.
As a consequence, on the one hand the 90\% least active users ($ACR$) {were} responsible for posting most of the mentions and retweets. On the other, the top $1\%$ most popular users ($PC1 \cup PC2$) {received} most of the mentions and retweets.

Also, $89.77\%$ of users from $PCR$ {were} among the  $90\%$ least active users, i.e., {they belonged} to $ACR$. 
 {A total of} $7.01\%$ of users {who} ranked between the $1\%$ {and the} $10\%$ most popular users (in $PC3$) are {also between} the $1\%$ {and} $10\%$ most active ones (in $AC3$). Only $1.18\%$ of the top $1\%$ most popular users  (in $PC1 \cup PC2$)  belong to the top $1\%$ most active users (in $AC1 \cup AC2$).
 {None of the 0.1\% most popular users, i.e. in $PC1$,  is among the top 0.1\% most active users, i.e. in $AC1$.}
 Since users from $PC1 \cup PC2$  are responsible for $82.37\%$ of the tweets, {it is possible to conclude} that the most popular users are not necessarily the most active {ones}. This conclusion makes sense since the top $0.1\%$ most popular users are news organizations or public figures, such as, for example, politicians and journalists, who get more retweets than effectively post tweets on Twitter.
The most popular users, therein, must effectively post popular tweets that receive most of the retweets and mentions.

Table \ref{tab:pop_tweets_in_users} exhibits the percentage of the tweets posted by users from each popularity category. {As can be seen in the} table, the higher the popularity {of the} tweets, the higher the popularity of the users who posted them.

\begin{table}[htb]
    \centering
    \caption{Percentage {of} tweets posted by users from each popularity category for network \textit{1st\_Round}.}
    \small
    \label{tab:pop_tweets_in_users}
    \setlength{\tabcolsep}{2pt}
    \begin{tabular}{cc|c|c|c|c|}
    \cline{3-6}
          &    & \multicolumn{4}{|c|}{Users} \\ \cline{3-6}
          &    & \multicolumn{1}{|c|}{$PCR$} & $PC3$ & $PC2$ & $PC1$ \\ \hline
\multicolumn{1}{|c}{\multirow{4}{*}{Tweets}} 
& \multicolumn{1}{|c|}{$TCR$} & 0.26\% & 94.88\% & 4.34\% & 0.52\%  \\\cline{2-6}
\multicolumn{1}{|c}{} & \multicolumn{1}{|c|}{$TC3$} &  0\% & 27.96\% & 67.68\% & 4.37\%\\ \cline{2-6}
\multicolumn{1}{|c}{} & \multicolumn{1}{|c|}{$TC2$} &0\% & 0\% & 44.16\% & 55.84\%  \\ \cline{2-6}
\multicolumn{1}{|c}{} & \multicolumn{1}{|c|}{$TC1$} & 0\% & 0\% & 0\% & 100\% \\ \hline
\end{tabular}
\end{table}

Table \ref{tab:network_validation_pop_act} shows an analysis of the activity and popularity of users {in} the networks introduced in this paper. In this table,   column  ``Network'' identifies the network under analysis; column  ``Contribution to tweets posted'' presents the percentage of mentions and retweets sent by the users from activity categories; column ``Contribution to {retweets}'' shows the percentage   of mentions and retweets received by the users from popularity categories; column ``Popular users ranked as active'' presents the percentage of users ranked into popularity categories who are also in activity categories; and column ``Popular tweets posted by users'' presents the percentage of the 90\% least and 10\% most popular tweets posted by the 90\%  and 10\%  most popular users, respectively.

\begin{table}[!h]
    \centering
    \scriptsize
    \caption{Analysis of popularity and activity of users {in} the proposed networks.}
    \label{tab:network_validation_pop_act}
    \setlength{\tabcolsep}{0.75pt}
    \begin{tabular}{|c||c|c|c|c||c|c|c|c||c|c|c||c|c|}
    \hline
        \multirow{5}{*}{Network}    & \multicolumn{4}{|c||}{\multirow{1}{*}{Contribution to}}  
        & \multicolumn{4}{|c||}{\multirow{1}{*}{Contribution to}}  
        & \multicolumn{3}{|c||}{\multirow{1}{*}{Popular users}} 
        & \multicolumn{2}{|c|}{\multirow{1}{*}{Popular tweets}} \\ 
        & \multicolumn{4}{|c||}{tweets posted}  
        & \multicolumn{4}{|c||}{retweets or mentions}  
        & \multicolumn{3}{|c||}{ranked as actives} 
        & \multicolumn{2}{|c|}{posted by users} \\ \cline{2-14}   
        &\multicolumn{1}{|c|}{\multirow{3}{*}{$ACR$}} & \multicolumn{1}{|c|}{\multirow{3}{*}{$AC3$}} & \multicolumn{1}{|c|}{\multirow{3}{*}{$AC2$}}&
        \multicolumn{1}{|c||}{\multirow{3}{*}{$AC1$}}   
        &\multicolumn{1}{|c|}{\multirow{3}{*}{$PCR$}} & \multicolumn{1}{|c|}{\multirow{3}{*}{$PC3$}}   & \multicolumn{1}{|c|}{\multirow{3}{*}{$PC2$}}&
        \multicolumn{1}{|c||}{\multirow{3}{*}{$PC1$}}
        & \multicolumn{1}{|c|}{$ACR$} & \multicolumn{1}{|c|}{$AC3$}  & \multicolumn{1}{|c||}{$AC1\cup AC2$} 
        & \multicolumn{1}{|c|}{$PCR$} & \multicolumn{1}{|c|}{$PC1\cup PC2\cup$ } 
        \\ 
        &  &  &  &  &  &   & &  
        & in  &  in &  in
        & in  &   $PC3$ in $TC1 \cup $\\ 
        &     &    &  &  &  &   & &  
        & $PCR$  & $PC3$ &  $PC1 \cup PC2$ 
        & $TCR$  &   $TC2 \cup TC3$ \\ \hline
\textit{Politicians} & 53.24\% & 30.66\% & 11.03\% & 5.07\% & 5.11\% & 29.15\% & 36.88\% & 28.87\% & 91.04\% & 15.08\% & 5.56\% & 44.59\% & 100\% \\ \hline
\textit{Lula} & 44.73\% & 35.61\% & 14.05\% & 5.62\% & 2.91\% & 25.86\% & 39.14\% & 32.09\% & 91.11\% & 15.03\% & 5.97\% & 90.34\% & 9.63\% \\ \hline
\textit{Bolsonaro} & 58.65\% & 29.89\% & 6.8\% & 4.65\% & 22.37\% & 34.37\% & 18.62\% & 24.64\% & 89.94\% & 5.13\% & 10\% & 85.62\% & 16.67\% \\ \hline
\textit{LavaJato} & 44.63\% & 35.76\% & 13.99\% & 5.61\% & 3.2\% & 26.86\% & 38.54\% & 31.39\% & 91.11\% & 15.23\% & 7.04\% & 91.29\% & 11.58\% \\ \hline
{\textit{1st\_Round}} & 61.08\% & 28.54\% & 8.28\% & 2.11\% & 1.88\% & 15.75\% & 29.62\% & 52.75\% & 89.77\% & 7.01\% & 1.18\% & 27.09\% & 100\% \\ \hline
{\textit{2nd\_Round}} & 61.53\% & 28.03\% & 8.31\% & 2.14\% & 0.87\% & 17.61\% & 35.43\% & 46.09\% & 89.5\% & 4.73\% & 1.39\% & 14.49\% & 100\% \\ \hline
\end{tabular}
    \label{tab:my_label}
\end{table}

The values in Table \ref{tab:network_validation_pop_act} corroborate with the findings on network \textit{1st\_Round}: the 90\% least active users ($PCR$) are responsible for posting most of the mentions and retweets; the top $1\%$ most popular users  receive most of the mentions and retweets; the top $1\%$ most popular users are mostly not ranked in the top $1\%$ most {active}; and less than $11\%$ of the $1\%$ to $10\%$ most {popular users} ($PC3$) are also ranked as the $1\%$ to $10\%$ most {active users} ($AC3$).

However, for all the networks but \textit{Politicians}, \textit{1st\_Round} and \textit{2nd\_Round}, the percentage of  the top $10\%$ most popular users ($PC1 \cup PC2 \cup PC3$) who posted the top $10\%$ most popular tweets ($TC1\cup TC2 \cup TC3$) {ranged} from $9.28\%$ to $16.67\%$. According to this result, the tweets that {received} most of the retweets were not necessarily posted by the most popular users{. The} users' popularity might {therefore be} due to: (i) retweets received from a plethora of tweets; or (ii) mentions received from other users.

In the next section, {like} \citet{Sluban2015}, {we shall} analyze the influence of the most popular users on the communities.

\subsection{Influence of the Most Popular Users and Tweets}

The popularity of a community is defined by the sum of the popularity of its users, that is, the total out-degree of the vertices that belong to the community.

\begin{center}        
        \begin{figure}[!htb]
            \includegraphics[width=1\textwidth]{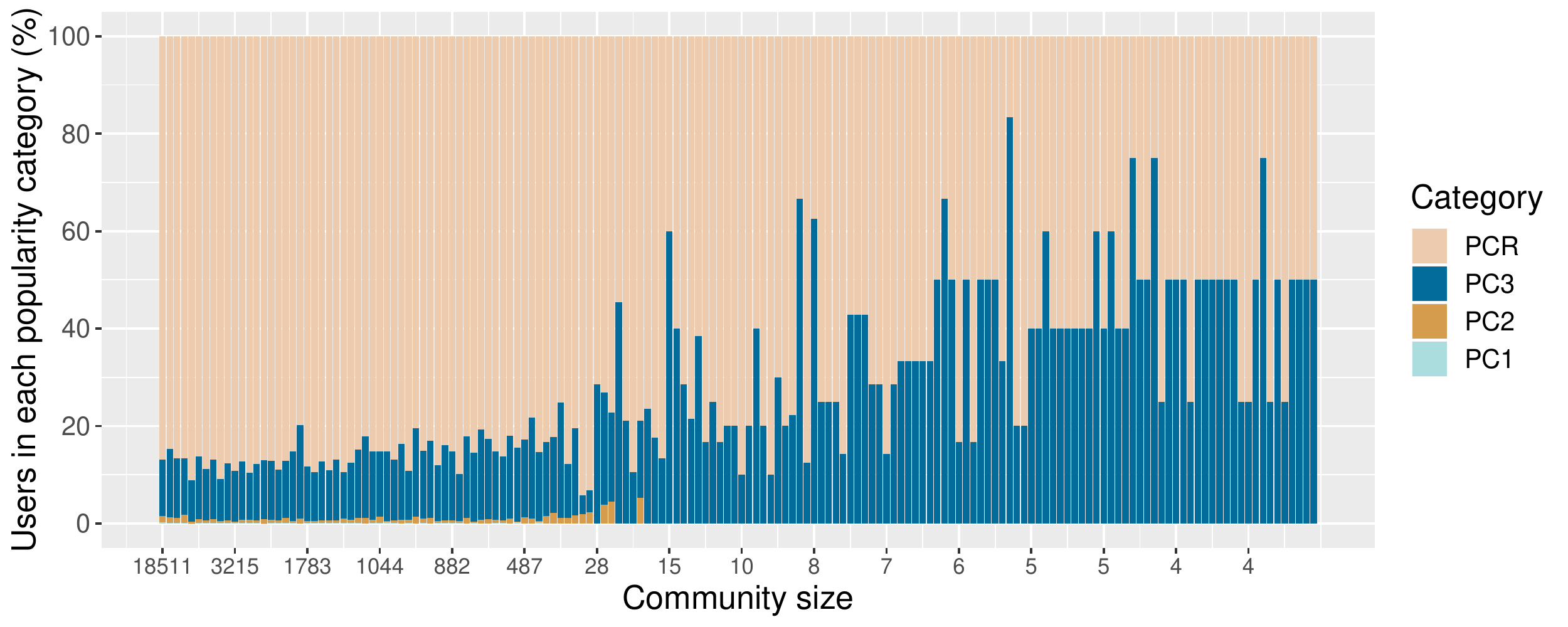}\caption{{Percentage of users in the communities found by Louvain method {in} network \textit{1st\_Round}.}}\label{fig:stream_part_comm_number}
    \end{figure}
\end{center}

\begin{center}        
        \begin{figure}[!htb]
            \includegraphics[width=1\textwidth]{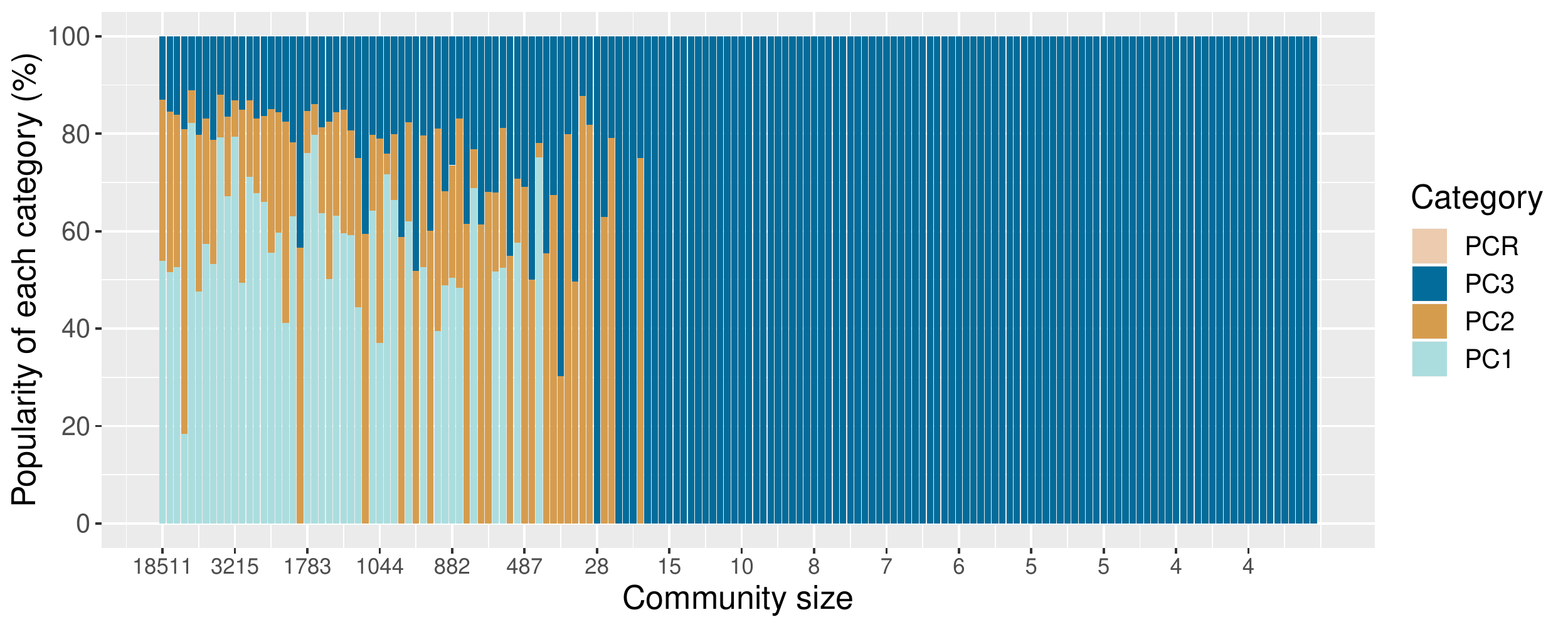}\caption{{Contribution of each popularity category, in percentage, to the popularity of the communities found by Louvain method in network \textit{1st\_Round}.}}\label{fig:stream_part_comm_pop}
    \end{figure}
\end{center}

Figure \ref{fig:stream_part_comm_number} presents the percentage of users from each popularity category {in} the communities of network \textit{1st\_Round}{. Each community is described on} the x-axis {according to} their size. Figure  \ref{fig:stream_part_comm_pop} 
 shows the contribution of the users from each popularity category{, in percentage,} to the total popularity of the communities found by Louvain in network \textit{1st\_Round}. {Again, t}he x-axis {indicates} the size of the communities, which is the number of vertices in a community.

Although in smaller number, the top $0.1\%$ most popular users ($PC1$) are responsible {for} the majority of the community's  popularity for most of the communities with more than 28 vertices. Users from $PC2$ {have also significantly contributed} to the {popularity of the communities.} As expected, although the $90\%$ least popular users ($PCR$) are the majority of users in the communities, they do not contribute to the popularity of the communities.

\begin{center}
    \begin{figure}[!htb]
        \begin{minipage}[t]{0.49\textwidth}
         \includegraphics[width=1\textwidth]{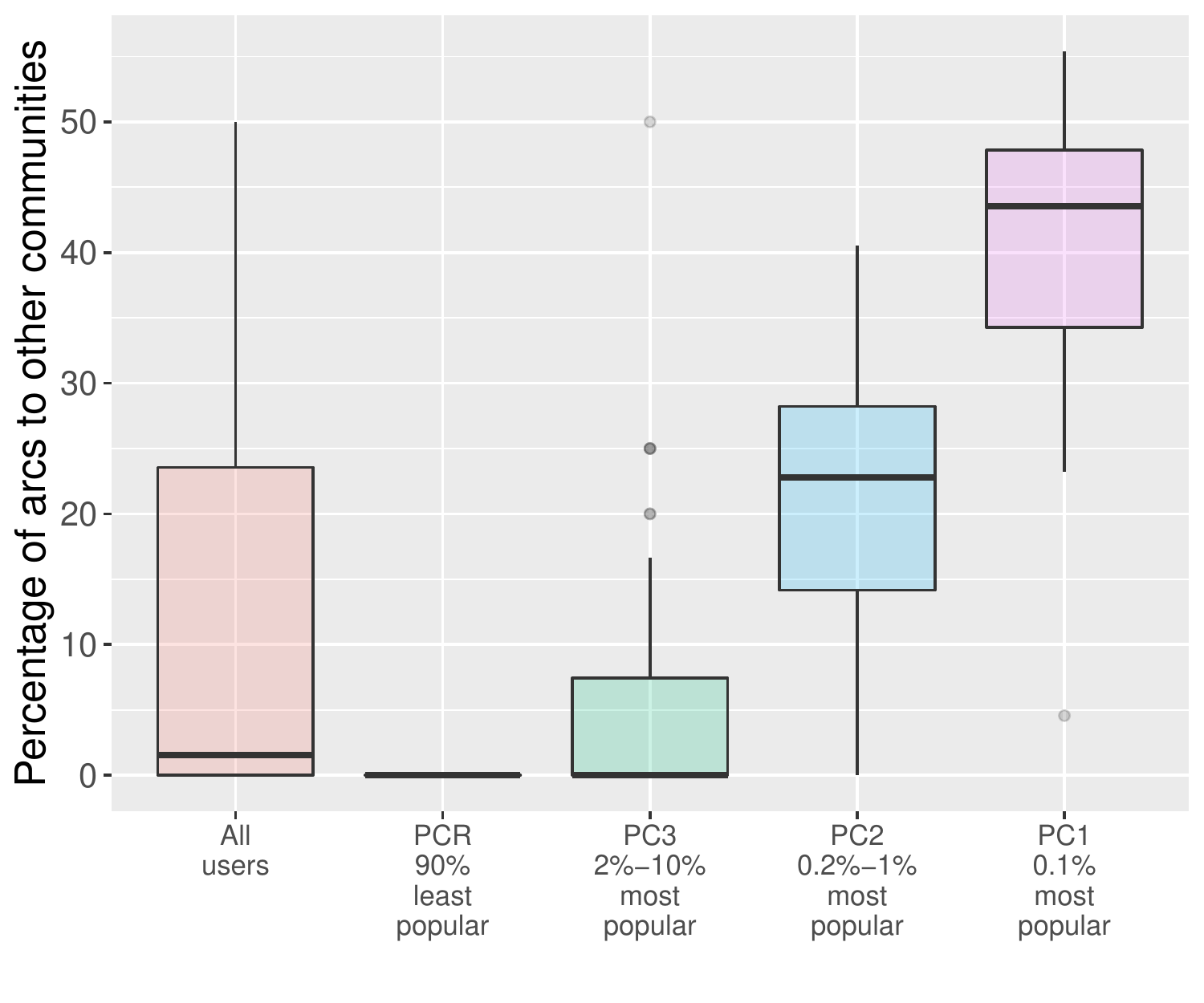}
               \caption{Influence exerted by users {of} the communities {in} network \textit{1st\_Round}.}
              \label{fig:stream_part_comm_influence}
        \end{minipage}
        \quad
            \begin{minipage}[t]{0.49\textwidth}
            \includegraphics[width=1\textwidth]{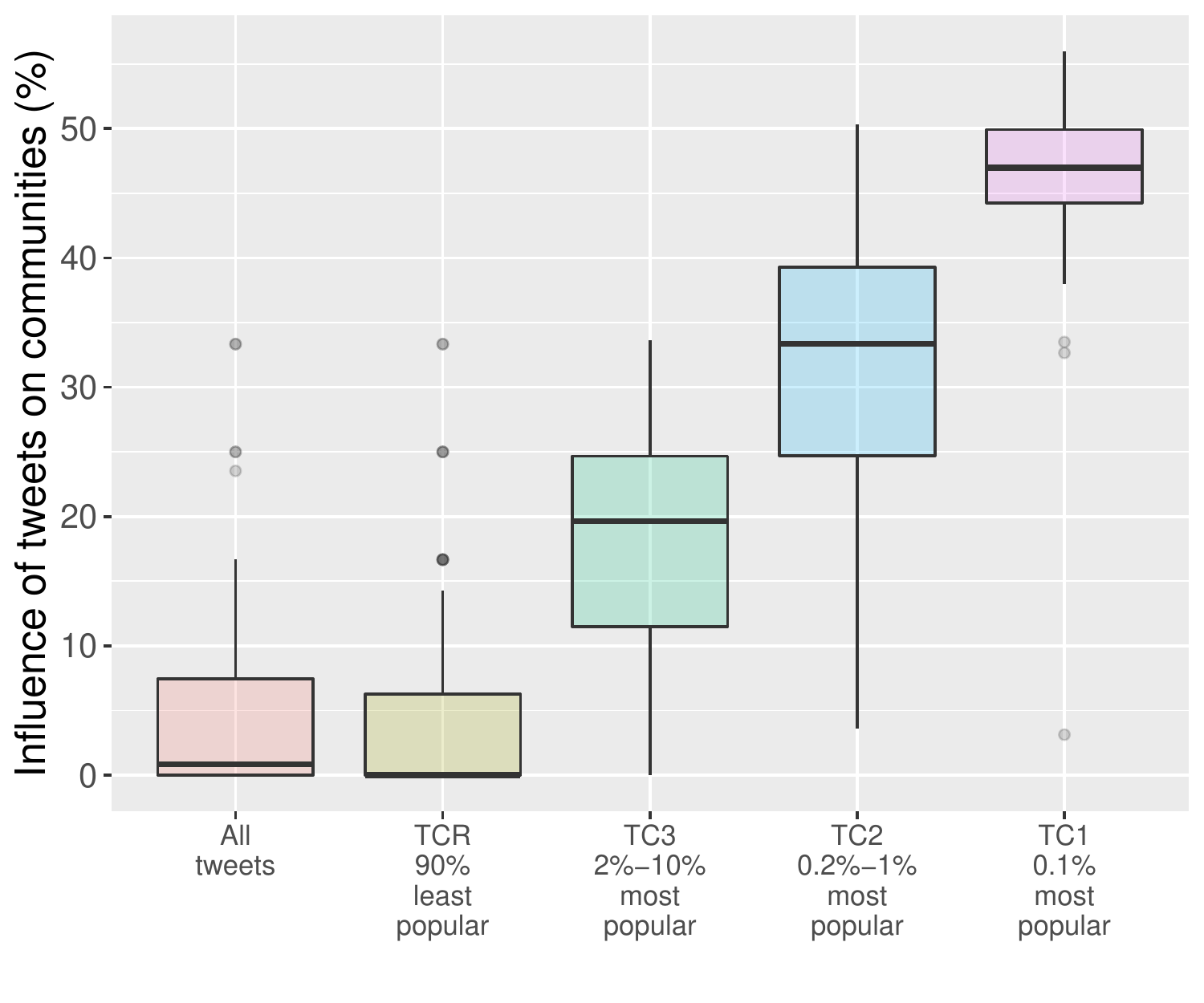}\caption{Influence exerted by tweets {of} the communities {in} network \textit{1st\_Round}.}\label{fig:stream_part_comm_influence_tweets}
        \end{minipage}
    \end{figure}
\end{center}

The social influence on a community is the percentage of the retweets or mentions its users receive {made by} users {from} other communities \citep{Sluban2015}. Figures \ref{fig:stream_part_comm_influence} and \ref{fig:stream_part_comm_influence_tweets} summarize the influence exerted by, respectively, users and tweets from each popularity category {in} the communities.

The $90\%$ least popular users have an average influence very close to $0$, {meaning that most of the retweets and mentions they receive are made by users within} their community. Users from $PC2$ and $PC1$ have higher social influence on their communities than $PC3$. A similar conclusion can be drawn on the influence of the most retweeted tweets on communities. The social influence concerning tweets from $TC3$, $TC2$ and $TC1$ are substantially higher than of users from $TCR$, which suggests that the $10\%$ most retweeted tweets are indeed good candidates to be viral tweets.

\begin{figure}[!htb]
    
\begin{center}
        \includegraphics[width=0.49\textwidth]{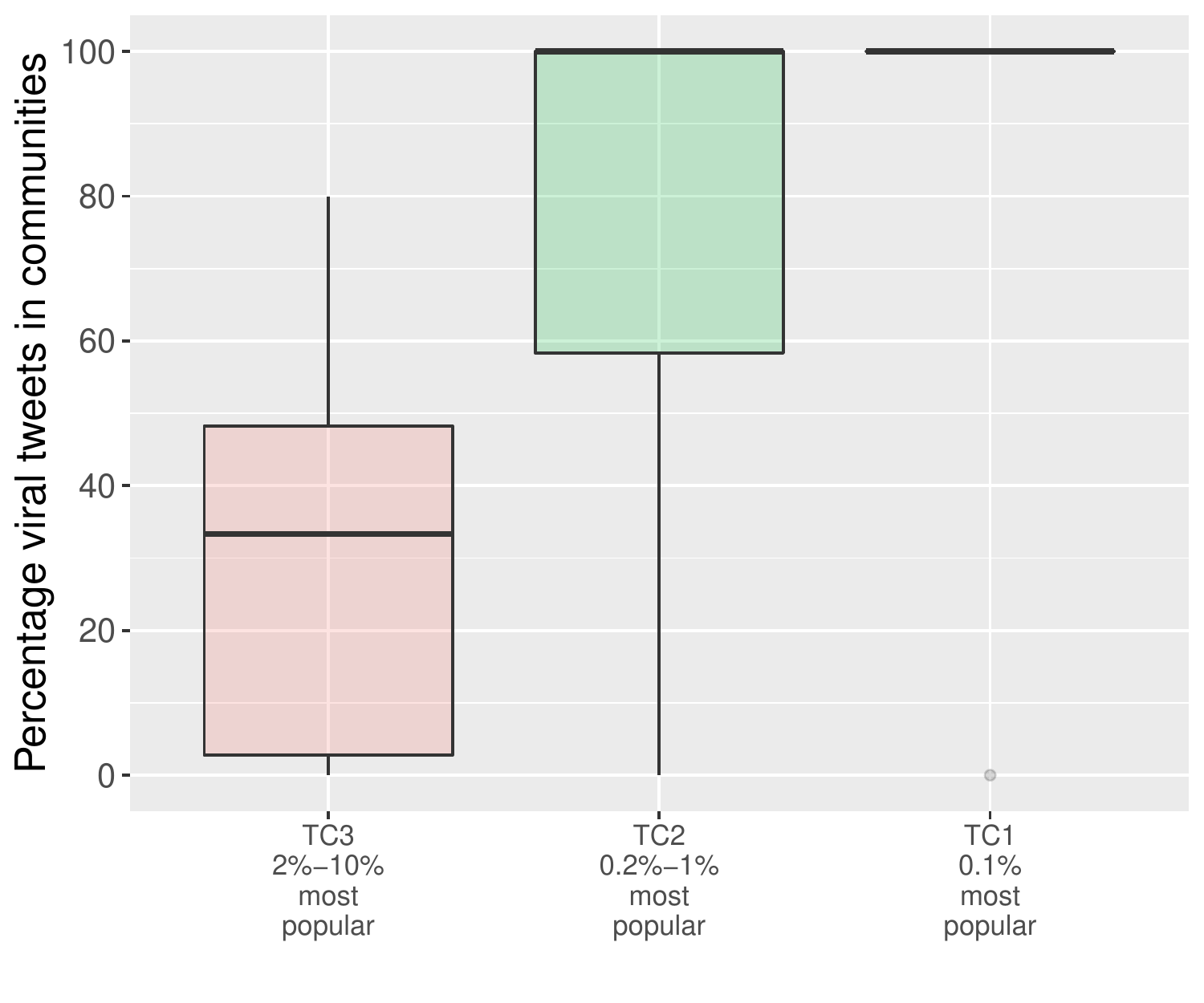}
        \caption{Percentage of viral tweets in the communities of network \textit{1st\_Round}.}
        \label{fig:comm_viral_Elections_1st}
        
\end{center}
    \end{figure}

We classify a tweet as viral if its virality is greater than the threshold $0.25$, that is, at least $25\%$ of the retweet it receives is from communities other than the one it {originated from}.
Figure \ref{fig:comm_viral_Elections_1st} exhibits the percentage of viral tweets in the communities of network \textit{1st\_Round} regarding each tweet popularity category.
According to this figure, the higher the popularity of the tweets, the higher {their} chance {of} being viral. In particular, all of the $0.1\%$ most popular tweets ($TC1$) are viral.

Table \ref{tab:validation_community} presents a brief analysis of the communities found by the Louvain method {in the} networks introduced in this paper. In this table, columns ``Number of users in communities (\%)'' and ``Popularity of users in communities (\%)'' exhibit the percentage of users from each popularity category and their popularity, respectively, {in} the communities. Column ``Viral tweets'' shows the percentage of the $10\%$ most popular tweets {that} are viral. {The columns} marked as ``avg'' and ``std'' exhibit, respectively, the average and standard deviations of the aforementioned values.

\begin{table}[htb]
    \centering
    \scriptsize
    \caption{Analysis of the communities found by Louvain for the constructed networks.}
    \label{tab:validation_community}
    \setlength{\tabcolsep}{1pt}
    \begin{tabular}{|c||c|c|c|c|c|c|c|c||c|c|c|c|c|c|c|c||c|c||c|c|}
    \hline
    \multicolumn{1}{|c||}{\multirow{3}{*}{Network}}  &  \multicolumn{8}{|c||}{Number of users in communities (\%)} &  \multicolumn{8}{|c||}{Popularity of users in communities (\%)} &  \multicolumn{2}{|c||}{Viral} & \multicolumn{2}{|c|}{Virality of}\\   \cline{2-17}
    & \multicolumn{2}{|c|}{$PCR$}   &  \multicolumn{2}{|c|}{$PC3$}   & \multicolumn{2}{|c|}{$PC2$}        & \multicolumn{2}{|c||}{$PC1$}    & \multicolumn{2}{|c|}{$PCR$}   &  \multicolumn{2}{|c|}{$PC3$}   & \multicolumn{2}{|c|}{$PC2$}        & \multicolumn{2}{|c||}{$PC1$} & \multicolumn{2}{|c||}{ tweets (\%)} & \multicolumn{2}{|c|}{ tweets (\%)}  \\  \cline{2-21}
                        & avg & std & avg & std & avg & std & avg & std 
                          & avg & std & avg & std & avg & std & avg & std & avg & std & avg & std \\ \hline
\textit{Politicians} & 88.85 & 15.46 & 10.75 & 15.39 & 0.38 & 1.09 & 0.02 & 0.07 &42.4 & 46.81 & 65.29 & 35.43 & 56.93 & 27.27 & 55.85 & 27.32 
& 32.78 & 31.41
& 10.14 & 10.78 
\\ \hline
\textit{Lula}  & 78.62 & 18.1 & 20.95 & 18.27 & 0.41 & 0.72 & 0.03 & 0.07 &11.6 & 22.96 & 65.69 & 34.08 & 53.07 & 24.22 & 54.05 & 29.54 
& 100 & 0   
& 91.11 & 11.36
  \\ \hline
\textit{Bolsonaro} & 87.58 & 9.16 & 11.67 & 9.7 & 0.64 & 1.12 & 0.11 & 0.37 &  31.45 & 19.94 & 49.03 & 25.99 & 62.2 & 26.66 & 75.37 & 16.26 & 
97.44 & 9.25  & 87.35 & 19.01
\\ \hline
\textit{LavaJato} & 80.2 & 14.16 & 19.45 & 14.31 & 0.33 & 0.68 & 0.02 & 0.07 &  17.77 & 29.22 & 67.76 & 32.26 & 50.2 & 25.04 & 51.54 & 31.1 
& 100 & 0  
& 92.55 & 9.36
\\ \hline
 {\textit{1st\_Round}} & 71.47 & 17.31 & 28.1 & 17.63 & 0.4 & 0.76 & 0.03 & 0.05 &  0 & 0 & 70.43 & 37.59 & 36.07 & 22.88 & 58.56 & 13.1 
 & 55.54 & 40.09
 & 4.6 & 6.33 \\ \hline
{\textit{2nd\_Round}} & 74.72 & 15.6 & 24.91 & 15.93 & 0.34 & 0.66 & 0.03 & 0.05  & 0 & 0 & 72.7 & 37.87 & 41.14 & 25.69 & 56.4 & 21.42  
& 47.24 & 41.86  
& 7.99 & 9.98 
\\ \hline

 \end{tabular}
\end{table}

On the one hand, the results of ``Number of users in communities (\%)'' and ``Popularity of users in communities (\%)'' presented in Table \ref{tab:validation_community} confirm the preceding analysis of network \textit{1st\_Round}: the most popular users are responsible for most of the popularity of the communities. On the other hand, all or almost all the $10\%$ most popular tweets are viral on the remaining networks.

{Except for the} \textit{Politicians}, \textit{1st\_Round} and \textit{2nd\_Round} {networks, tweets had} viralities higher than $83.75\%$.  The average virality of tweets from \textit{Politicians}, \textit{1st\_Round} and \textit{2nd\_Round} networks, on the other hand, {were} lower than $10.14\%$.

Next section {presents} an analysis {of} the growing patterns {in} network \textit{1st\_Round}.

\subsection{Temporal Analysis of \textit{1st\_Round} Network}\label{subsec:temporalanalysis}

\begin{center}
    \begin{figure}[!htb]
        \begin{minipage}[t]{0.49\textwidth}
         \includegraphics[width=1\textwidth]{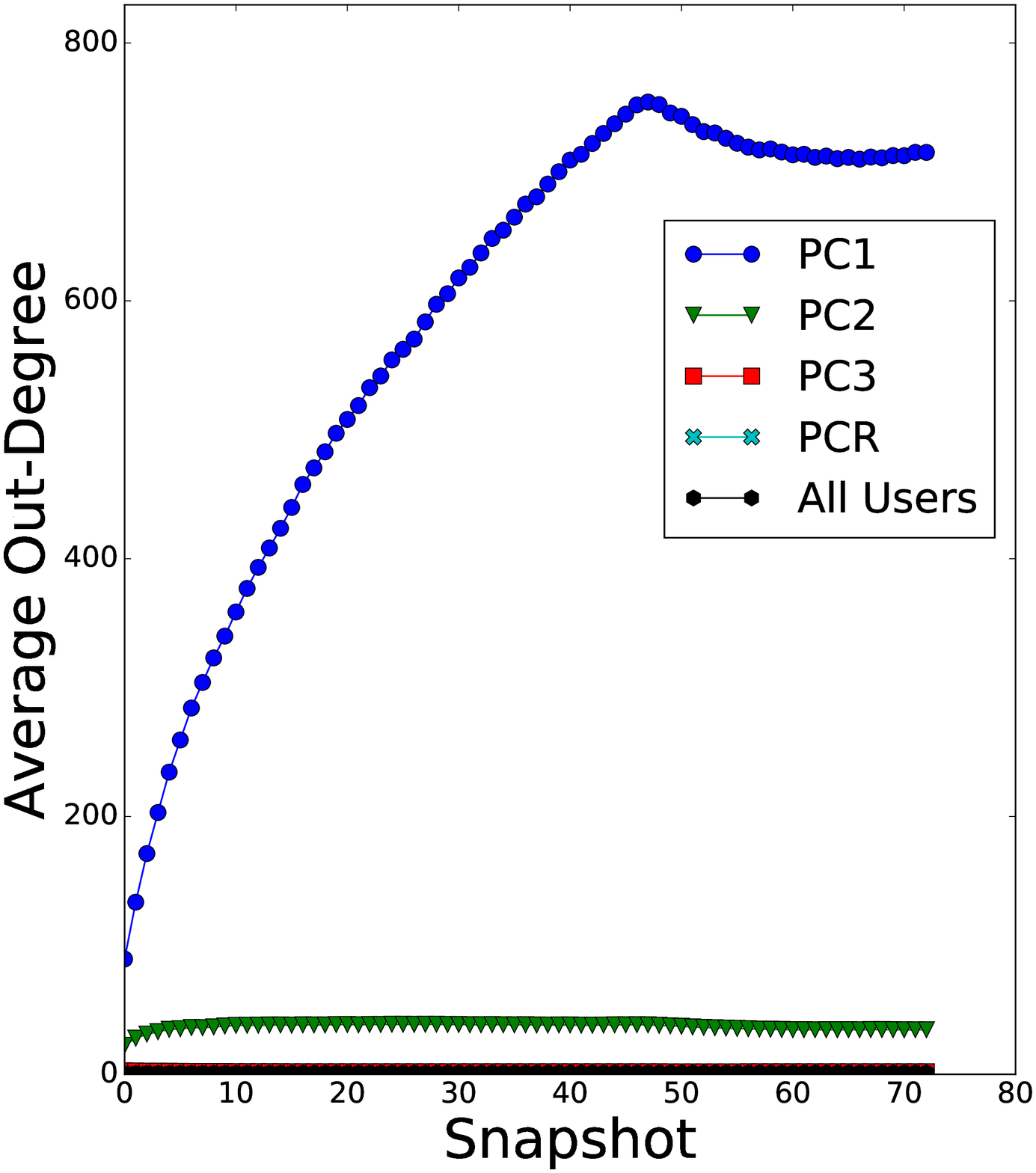}
              \caption{Average out-degree of users {in} popularity categories along time windows.}
              \label{fig:temporal_outdegree}
        \end{minipage}
        \quad
            \begin{minipage}[t]{0.49\textwidth}
            \includegraphics[width=1\textwidth]{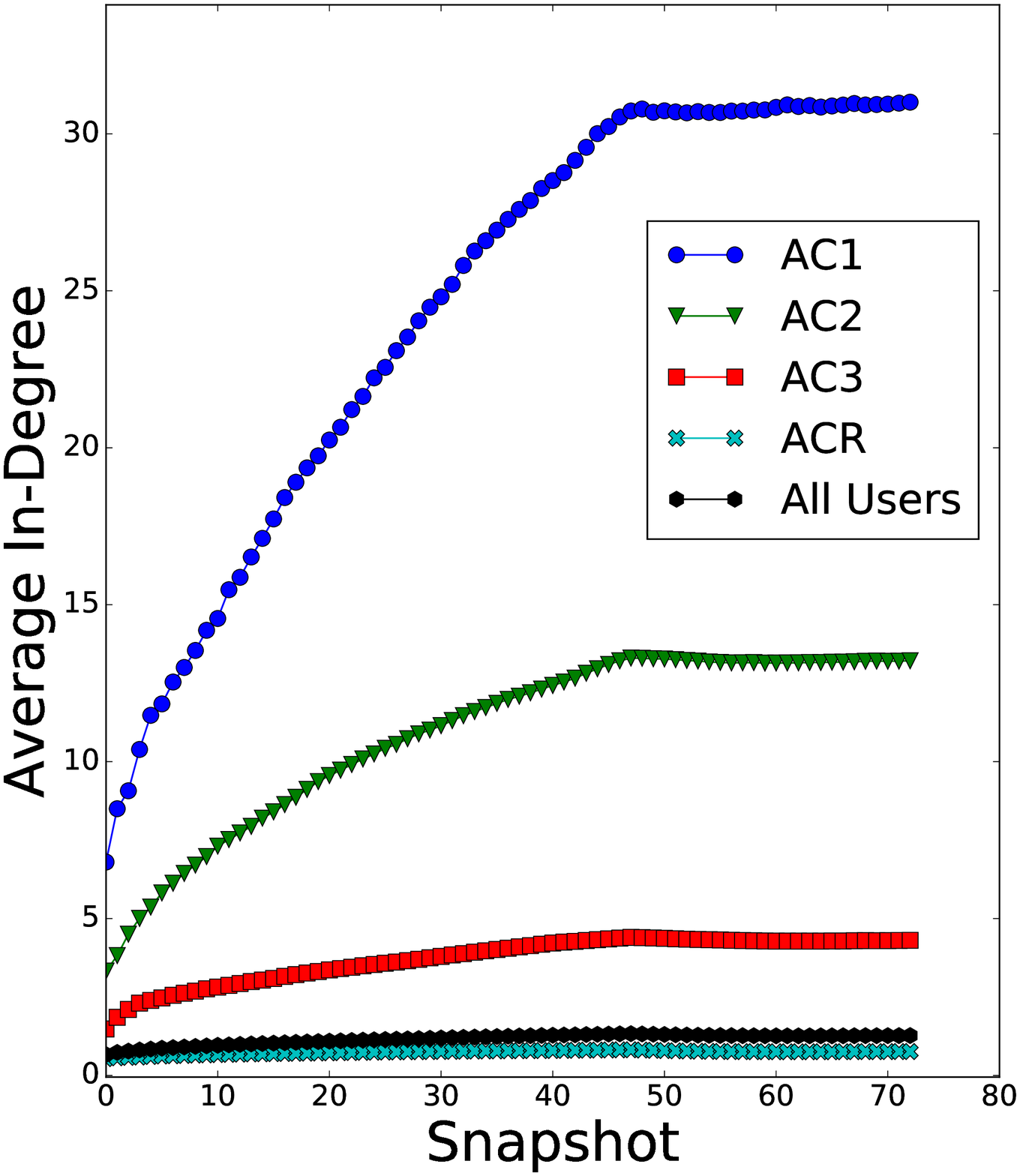}\caption{Average in-degree of users {in} activity categories along time windows.}\label{fig:temporal_indegree}
        \end{minipage}
    \end{figure}
\end{center}

To understand how the popularity and activity of users from network  \textit{1st\_Round} evolves along time, we took snapshots of the {network} at every 5000 tweets. Each snapshot represents the network constructed from the tweets posted up to that time.
Figure \ref{fig:temporal_outdegree} presents the average out-degree, that is, the number of retweets received from users from each popularity set with regard to the snapshots. Figure \ref{fig:temporal_indegree} presents the average in-degree, that is, the number of mentions and retweets made by users from each activity set with regard to the snapshots.

The top $0.1\%$ most popular users ($PC1$) received an increasing number of retweets along time, as shown by their augmenting average out-degree.
This observation corroborates with the  remark that new vertices are more likely to connect with the existing vertices with the highest degrees in scale-free networks.

\begin{figure}[!htb]
\begin{center}
\includegraphics[height=0.5\columnwidth]{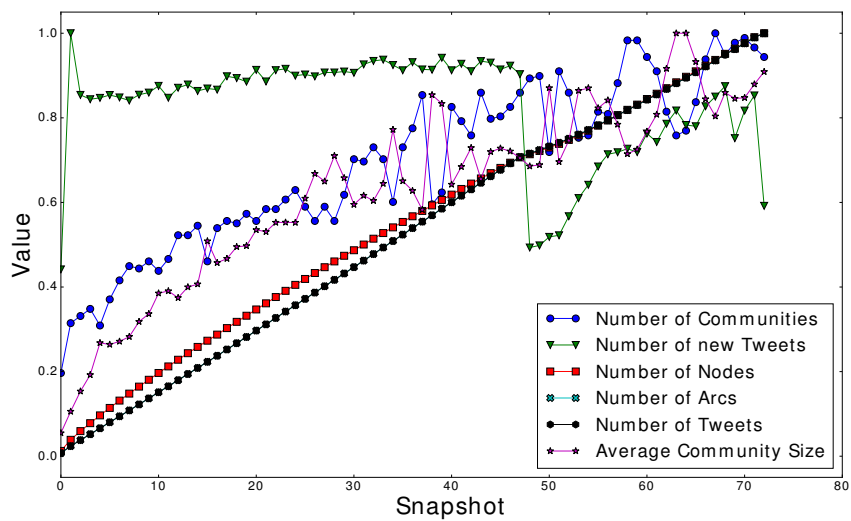}
\caption{Evolution of network \textit{Politicians} over time.} 
\label{fig:politiciansevolution}
\end{center}
\end{figure}

Figure \ref{fig:politiciansevolution} presents the evolution of network \textit{1st\_Round} along time. In each snapshot, the following are considered: the total number of communities in the network obtained by the Louvain method, the average community size, the total number of arcs and nodes, the number of new tweets and the total number of different tweets in the network. As shown in Table \ref{tab:network_network}, the network \textit{1st\_Round} has 135865 nodes and 242679 arcs. It is possible to notice in Figure \ref{fig:politiciansevolution} that the number of new tweets that appear in the network at each snapshot is {nearly} constant. However, at snapshot 48, the number of new tweets drops significantly and {then immediately starts to raise again.}

On average, 15\% of the new arcs that appear in the network in a given snapshot represent retweets and mentions of new tweets. The other 85\% of the new arcs represent new retweets and mentions of tweets {already} in the network in previous snapshots. The number of arcs created from a tweet decreases rapidly over time. In Figure \ref{fig:temporal_outdegree} it is possible to note {that the} popularity of users from ${PC1}$ {ceases} to increase just in snapshot 48, where the number of new tweets is smaller. 

\begin{figure}[!htb]
\begin{center}
\includegraphics[height=0.5\columnwidth]{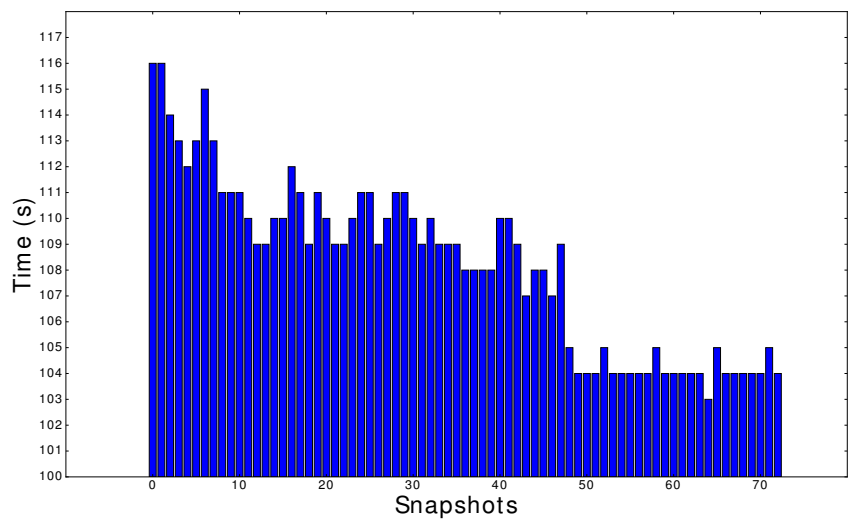}
\caption{Duration of each snapshot in seconds.}
\label{fig:snapshotduration}
\end{center}
\end{figure}

Figure \ref{fig:snapshotduration} shows the duration (in seconds) of each snapshot. The average duration of each snapshot was 108 seconds, which means that 5000 tweets were collected at every 108 seconds. The average duration of snapshots before and after snapshot 48 was, respectively, 110 and 104 seconds. {S}napshot 48 was collected {upon} the {disclosure} of the results of the first round {of the} elections {at} 19:00:00 (Brasília Time (BRT)), while snapshot {49} was collected from 19:00:38 to 19:02:23 (Brasília Time (BRT)). On the one hand, the average duration of the snapshots obtained after the release of the first {round of} results decreased. That is, the 5000 tweets collected in each snapshot were obtained in a shorter time window, which meant that Twitter users were more active. On the other hand, the number of new tweets {appearing} in snapshot 49 is smaller than in previous snapshots. This possibly means that, during the {disclosure} {of the} election results, users {created} fewer new tweets and interacted more with {previously posted tweets}. However, the rate of new tweets soars after snapshot 48, whose content was probably about the released results.

\section{Context and Sentiment {of the} Communities {in Relation} to the Elections}
\label{sec:community_classification}

The output of a community detection algorithm provides a straightforward classification of users into communities. Tweets can be classified according to the communities of the users who posted them. 
In this section the context of the communities and {their} sentiments {in relation}  to the words {in} the tweets {of} networks \textit{1st\_Round} and \textit{2nd\_Round} are studied.

Prior to the sentiment analysis, we first selected the words which were used at least 10 {times in the communities with a minimum of 1000 vertices}. Then, we translated all the words {into} English. Finally, we employed the R library \textit{sentimentr} \citep{sentimentr} to analyse the sentiment {related} {to} the selected words in each community. Each word is associated with a score in the range [-1,1], where -1 and 1 indicate the most negative and positive sentiments, respectively.

\begin{figure}[!htb]
\begin{center}
  \subfigure[fig][Word Cloud for network \textit{1st\_Round}.]{\label{fig:stream_wordCloud_Elections1}\includegraphics[width=0.49\textwidth]{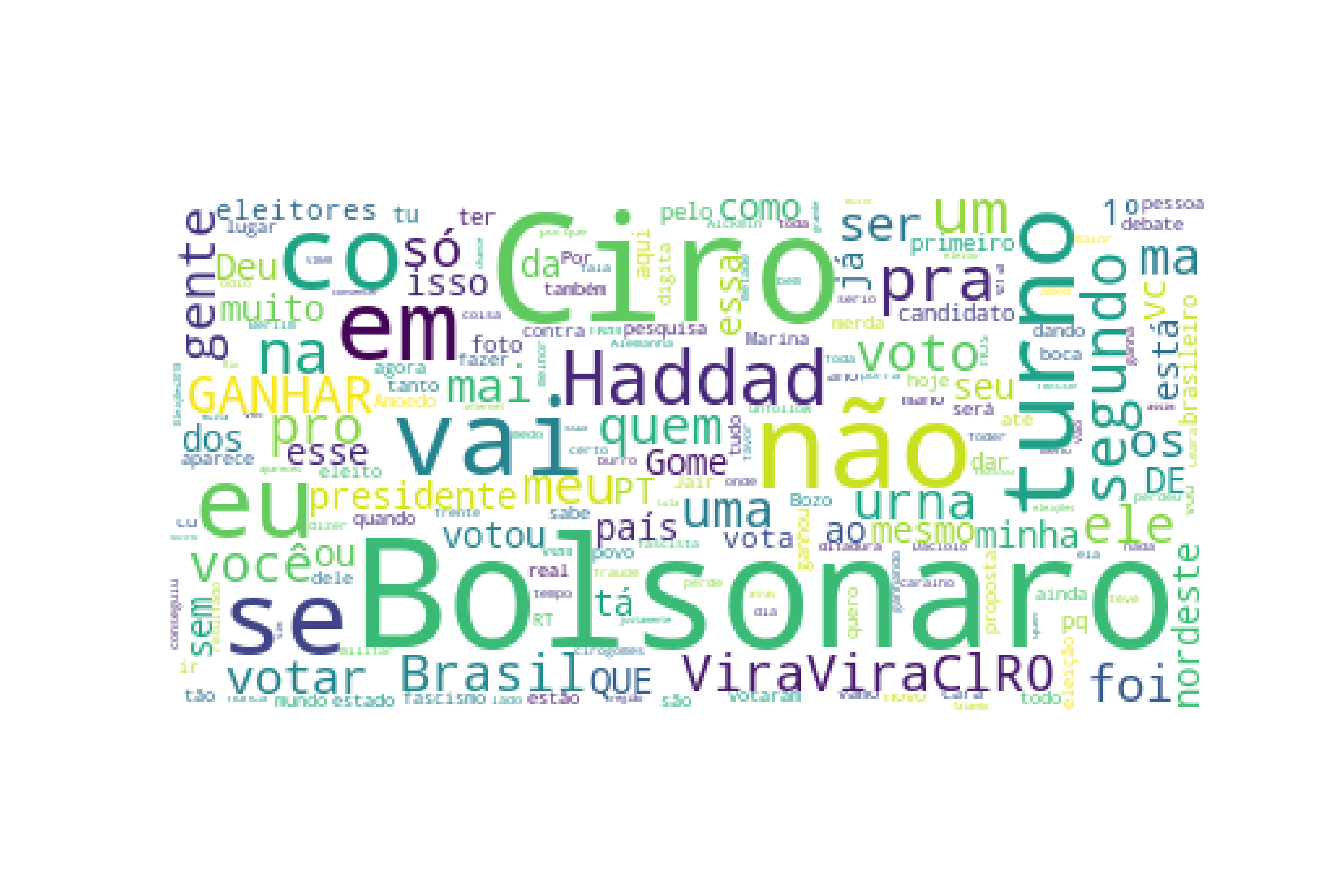}}
 \subfigure[fig][Word Cloud for network \textit{2nd\_Round}.]{\label{fig:stream_wordCloud_Elections2}\includegraphics[width=0.49\textwidth]{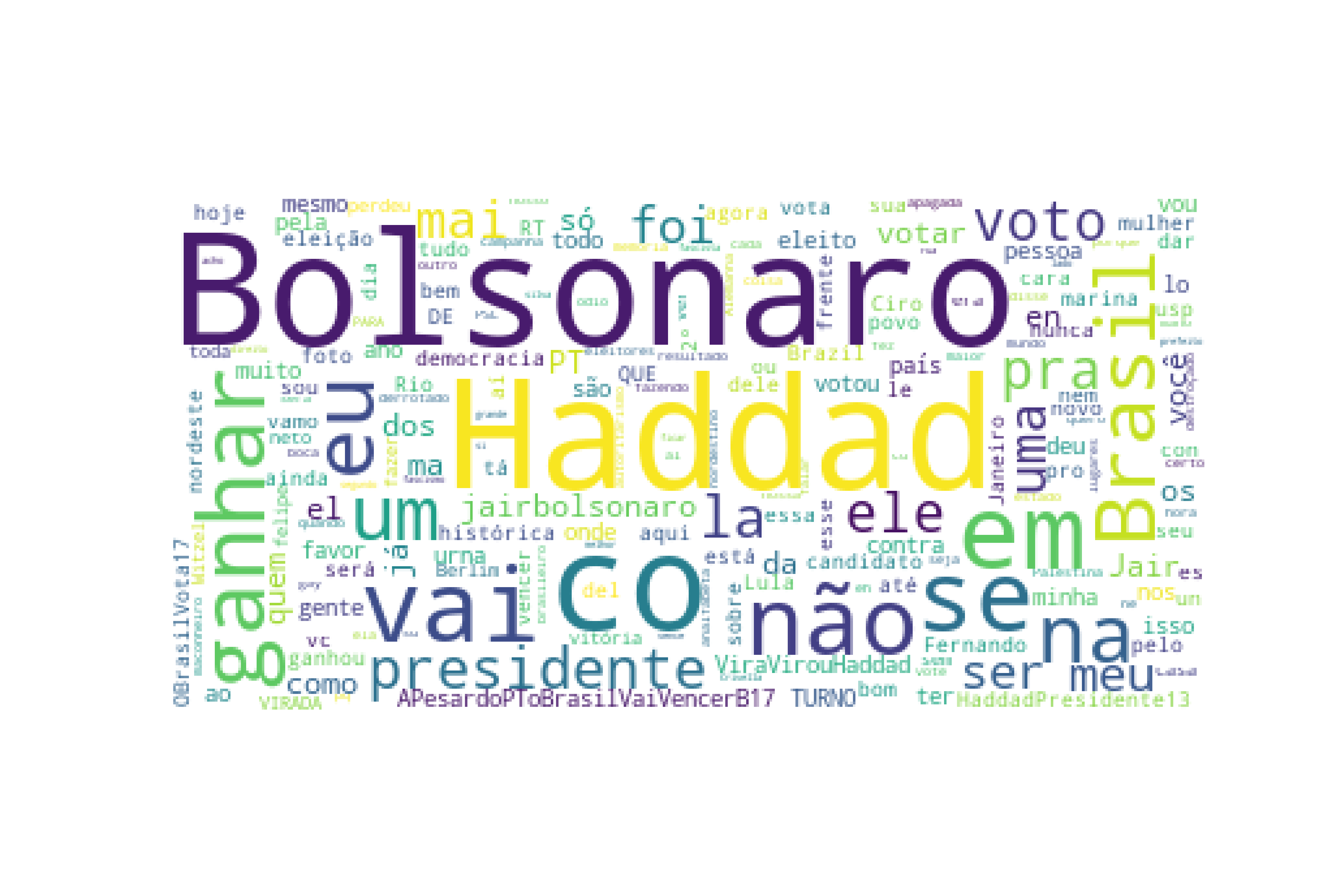}}
 \caption{Word clouds with the most frequent words.}
      \label{fig:word_clouds}
\end{center}
\end{figure}

Figures \ref{fig:stream_wordCloud_Elections1} and \ref{fig:stream_wordCloud_Elections2} illustrate word clouds with the most frequent words in the tweets from networks \textit{1st\_Round} and \textit{2nd\_Round}. The word cloud {in} Figure \ref{fig:stream_wordCloud_Elections1}  shows that the names of {the} {three most} voted candidates in the first round of the elections were {also amongst} the most used words. This word cloud also shows the words ``segundo'' and ``turno'', which  mean ``second'' and ``round'' in English, respectively, {suggesting} interest in the second round during the first round of the elections.  The word cloud {in} Figure \ref{fig:stream_wordCloud_Elections2} shows the name of the two candidates who ran {in} the second round of the elections.

\begin{figure}[!htb]
\begin{center}
  \subfigure[fig][Network \textit{1st\_Round}.]{\label{fig:stream_scatterPlot_Elections1}\includegraphics[width=0.49\textwidth]{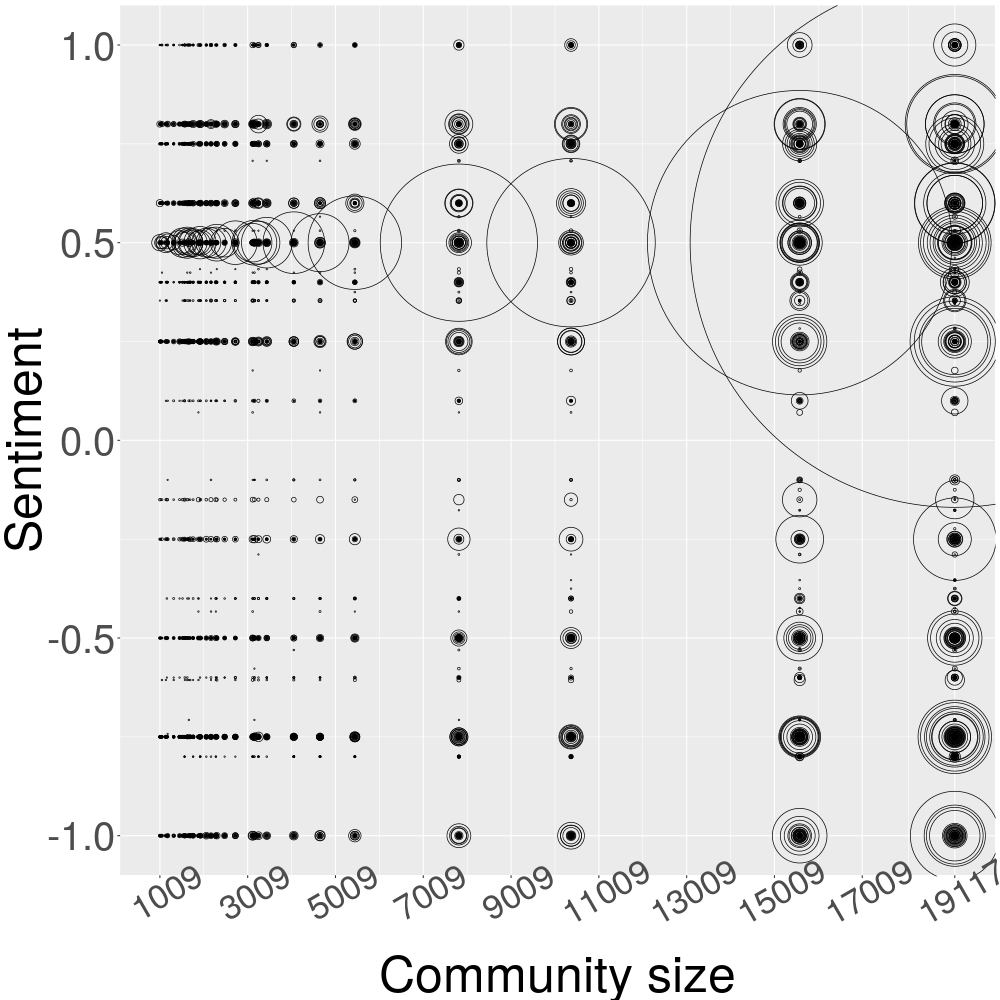}}
 \subfigure[fig][Network \textit{2nd\_Round}.]{\label{fig:stream_scatterPlot_Elections2}\includegraphics[width=0.49\textwidth]{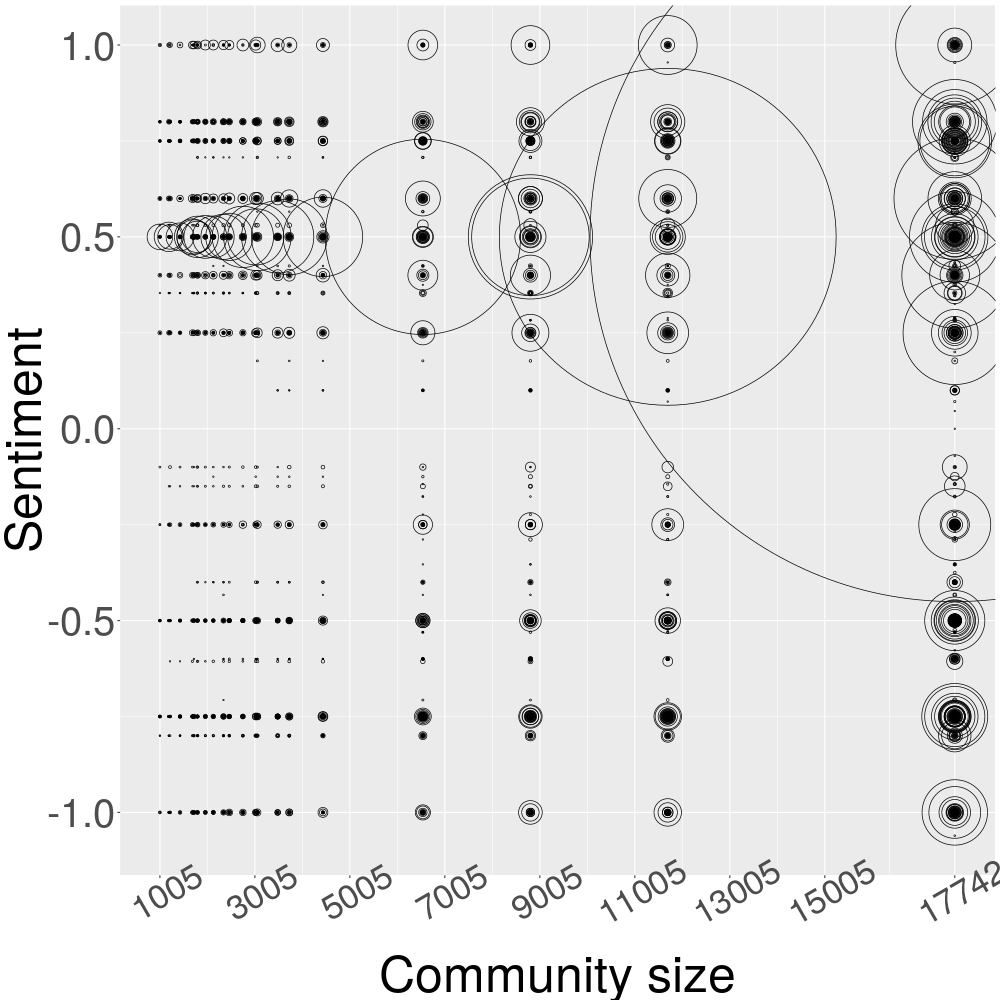}}
 \caption{Sentiments associated with the most common words on the communities.}
      \label{fig:scatterPlot_sentiment}
\end{center}
\end{figure}

Figures \ref{fig:stream_scatterPlot_Elections1} and \ref{fig:stream_scatterPlot_Elections2} show the sentiment of the most common words  in the tweets from networks \textit{1st\_Round} and \textit{2nd\_Round}. {The} x-axis {in the figures presents} the size of the communities, i.e., their number of vertices. The y-axis shows the sentiment of the words. The center of the circumferences indicates the sentiment of the words whereas their diameter indicates the frequency {of} the data collected: higher diameters {indicate words used more often}.
These figures show that the sentiment and frequencies of the words in the data collected regarding the first and second rounds of the elections are similar. The higher the size of the community, the higher the frequency of the words. Moreover, although the number of words associated with negative and positive sentiments is similar, the words associated with positive sentiments appear more often.

Based on the observation that the words associated with positive sentiments are more frequent, the sentiment of the words that appear more times in the tweets should {weigh more} in a general sentiment score. Thereby, we calculate the average sentiment score of each community as the sum of the {sentiments} associated with each word multiplied by its frequency in the community.

\begin{figure}[!htb]
\begin{center}
\includegraphics[width=0.90\textwidth]{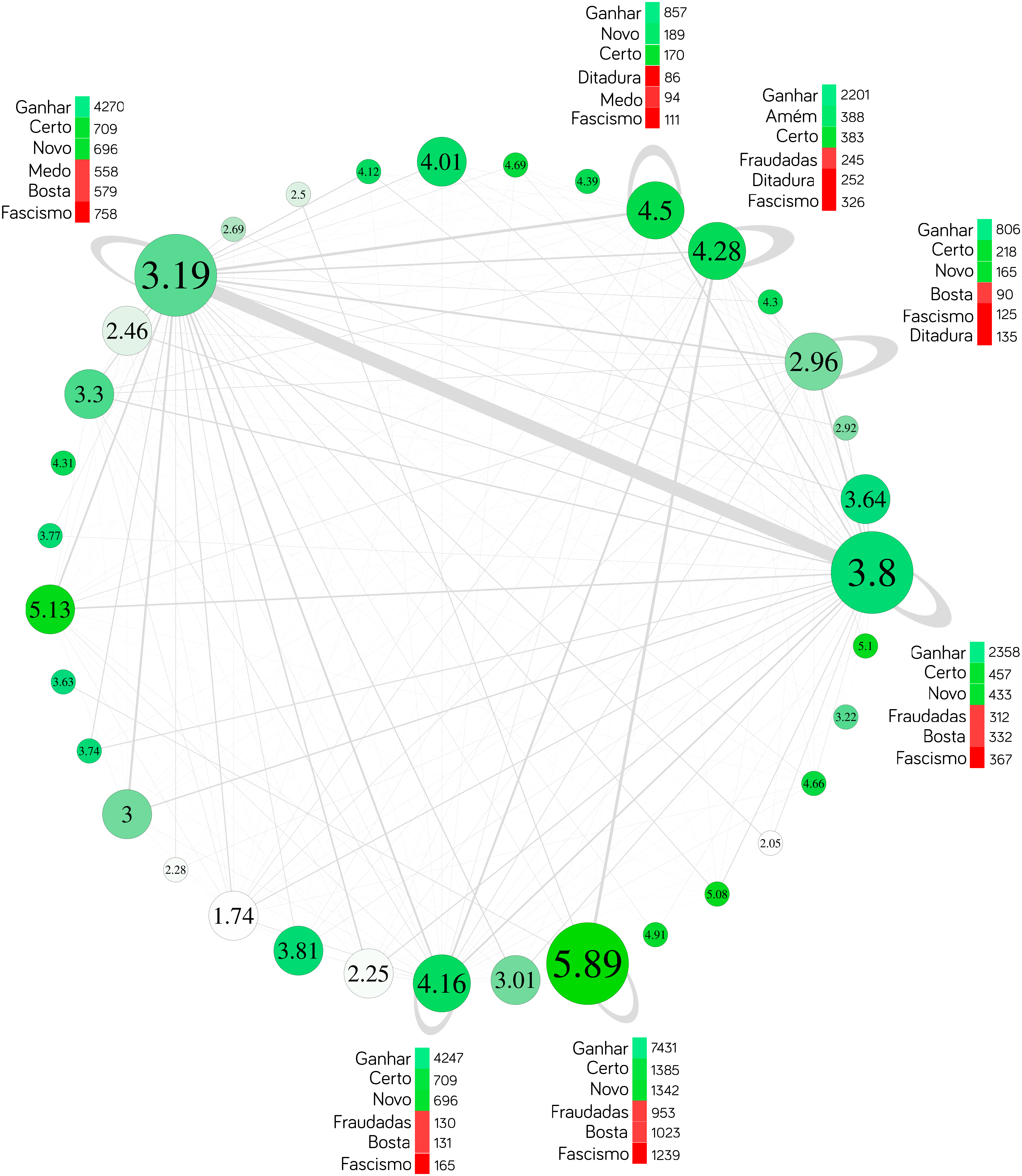}
  \caption{Average sentiment score and words associated with the highest positive and negative scores in communities considering network \textit{1st\_Round}.} \label{fig:stream_sentiment_Elections1}
  \end{center}
\end{figure}
\begin{figure}[!htb]
\begin{center}
\includegraphics[width=0.99\textwidth]{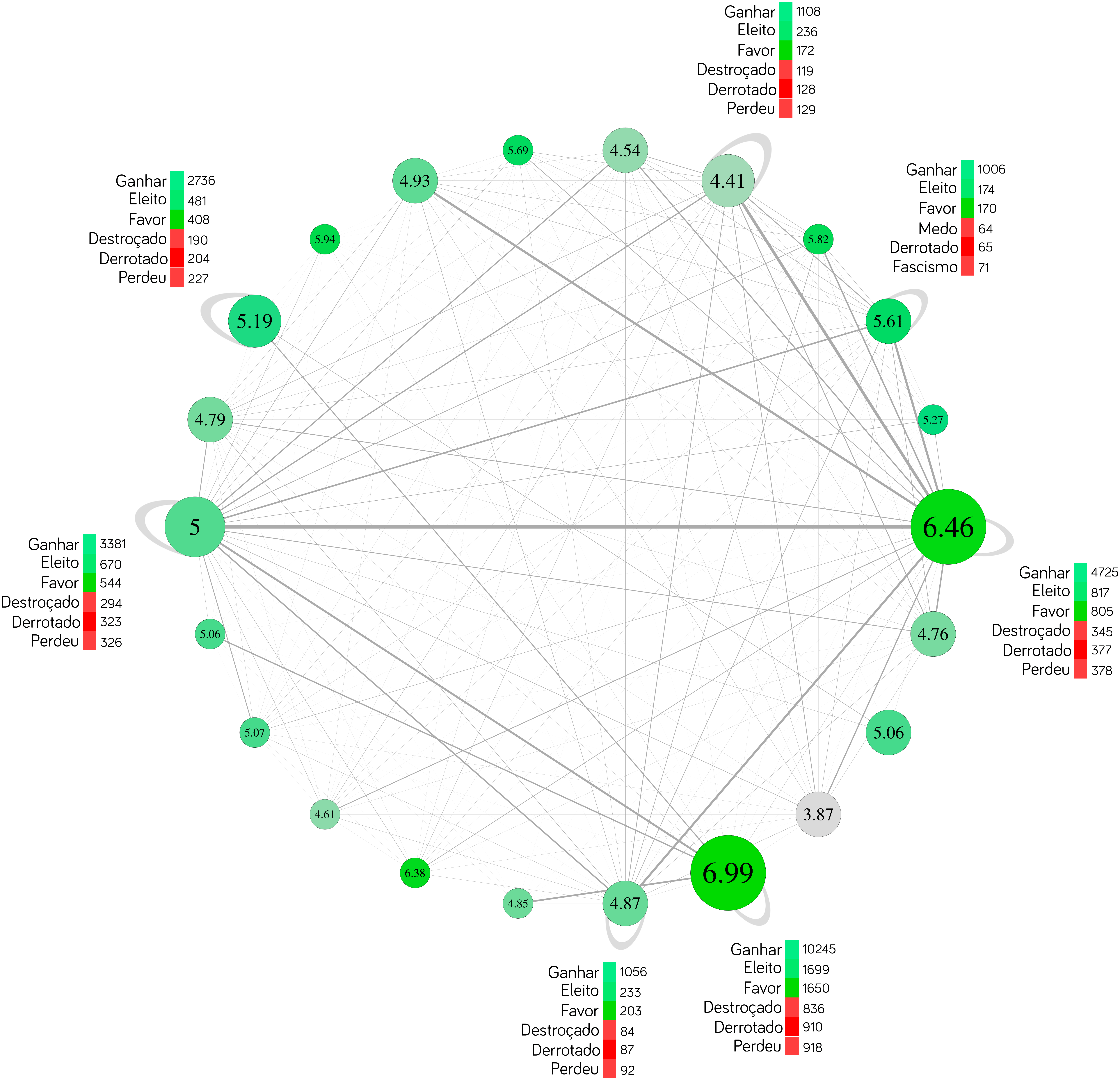}
  \caption{Average sentiment score and words associated with the highest positive and negative scores in communities considering network \textit{2nd\_Round}.} \label{fig:stream_sentiment_Elections2}
\end{center}
\end{figure}

Figures \ref{fig:stream_sentiment_Elections1} and \ref{fig:stream_sentiment_Elections2}
  present the average sentiment score of the communities with at least 1000 vertices. In these {figures}, the vertices belonging to the community found using the Louvain method are collapsed into supernodes and the edges between them represent the existence of arcs between its vertices. The larger the size of a supernode, the higher the number of vertices in the community. Similarly, the thickness of the edges {is} proportional to the amount of arcs between the communities.
  Additionally, for the 7 largest communities,  these figures also show the three negative and positive words associated with {the highest} and lowest scores, respectively. The word in Portuguese is presented {next to} {a} green or red square {whose color intensity is} proportional to the sentiment score and the number of times the word {was} used in the community.

The scores presented in Figures \ref{fig:stream_sentiment_Elections1} and \ref{fig:stream_sentiment_Elections2} show that, on average, the sentiment associated with the words in the communities was positive. However, this result is due to the higher frequency of positive words. These words are also amongst the most frequent words exhibited in the word clouds of Figures \ref{fig:stream_wordCloud_Elections1} and \ref{fig:stream_wordCloud_Elections2}.

Moreover, the words associated with the highest scores are the same in all the communities in both Figures \ref{fig:stream_sentiment_Elections1} and \ref{fig:stream_sentiment_Elections2}. These words occur in tweets classified as $TC1$, i.e, {the} $0.1\%$ most popular tweets, which according to Figure \ref{fig:comm_viral_Elections_1st}, are viral tweets. This observation corroborates with \citet{Wang2013}, who state that viral tweets spread across the communities.

\section{Final Remarks and Future Works}\label{sec:conclusion}

This paper {presents a thorough analysis of} the structure and topology of networks constructed from political discussions on Twitter and, in particular, {discusses} networks composed by tweets expressing reactions to the 2018 Brazilian presidential elections.

These networks are scale-free and {a} minority of their users are responsible for receiving {most} ``retweets'' and ``mentions'', while effectively posting few tweets. The evolution of the networks {along} time showed that new users are more likely to interact with the most popular users. We {verified} that the study {by} \citet{Weng2013} on the spreading patterns of hashtags {was} valid for the spreading of retweets on a political network regarding the 2018 Brazilian presidential elections: {most tweets were retweeted by users from within} the community of the user who originally posted the tweet but the minority of tweets is {retweeted} from several different communities and spread like viruses.

The most popular users, although in smaller number, dictate the popularity of the communities and the higher their popularity, the greater their influence on other communities. These users are also responsible for posting viral tweets.

Moreover, we studied the sentiment related to the most frequent words in the tweets with content about the 2018 Brazilian presidential elections and discovered that, overall, words associated with positive sentiments are predominant. The words that contribute most to the positive sentiment of the communities are the same and appear in viral tweets.

As a future application of this work, we intend to further investigate and classify tweets from larger databases, in particular, reactions to the 2018 Truckers' Strike and to the 2018 World Cup. To deal with the large scale networks, we plan to design distributed algorithms.
In addition, the study of the evolution of the networks with respect to spreading patterns of the tweets might help us to understand how communities are formed and the proliferation of tweets along time.
Finally, the observation that viral tweets do not affect the community structure of the networks and that non-viral tweets {get} trapped within the communities could be explored through community detection algorithms in order to provide better communities.

\section*{Acknowledgments}

The authors would like to acknowledge the financial support provided  by São Paulo Research Foundation (FAPESP): Grant Numbers 2016/22688-2; 2017/24185-0 and 2015/21660-4; Conselho Nacional de Desenvolvimento Científico e Tecnológico (CNPq): Grant Number 306036/2018-5; and Coordenação de Aperfeiçoamento de Pessoal de Nível Superior - Brasil (CAPES) - Finance Code 001. The authors are grateful to Professor Ana Carolina Lorena (Instituto Tecnológico de Aeronáutica) for providing a thoughtful review of the first draft of this paper. The last author would also like to thank Leonardo V. Rosset for giving her a hand.


\bibliography{bibliography}

\end{document}